\documentclass{article}

\usepackage{arxiv}
\usepackage{lmodern}
\usepackage{mathtools}
\usepackage[utf8]{inputenc}
\usepackage{verbatim}

\usepackage[english,french]{babel}
\usepackage{latexsym}
\usepackage[normalem]{ulem}
\usepackage{soul}

\usepackage{amsmath,amsthm,amsfonts,amssymb,empheq,relsize,alltt}
\usepackage{xparse}
\usepackage{algorithm}
\usepackage{algpseudocode}
\usepackage{bbm}
\DeclareMathAlphabet{\mathbbm}{U}{bbm}{m}{n}
\usepackage[singlelinecheck=0,font={sf,small},labelfont=bf]{caption, subfig}
\usepackage{cancel}
\usepackage{circuitikz}
\usepackage{tikz}
\usetikzlibrary{fit,positioning,arrows.meta,shapes}

\PassOptionsToPackage{colorlinks=true,allcolors=true}{hyperref}
\usepackage[noabbrev]{cleveref} 
\let\chyperref\cref 
\renewcommand{\cref}[1]{\hyperlink{#1}{\chyperref{#1}}} 

\usepackage{pdfpages} 
\usepackage{xstring}

\usepackage{circuitikz}
\usepackage{tikz}
\usetikzlibrary{fit,positioning,arrows.meta,shapes}

\usepackage{listofitems} 
\usetikzlibrary{arrows.meta} 
\usepackage[outline]{contour} 
\contourlength{1.4pt}
\tikzset{>=latex} 
\usepackage{xcolor}
\colorlet{myred}{red!80!black}
\colorlet{myblue}{blue!80!black}
\colorlet{mygreen}{green!60!black}
\colorlet{myorange}{orange!70!red!60!black}
\colorlet{mydarkred}{red!30!black}
\colorlet{mydarkblue}{blue!40!black}
\colorlet{mydarkgreen}{green!30!black}
\tikzstyle{node}=[thick,circle,draw=myblue,minimum size=22,inner sep=0.5,outer sep=0.6]
\tikzstyle{node in}=[node,green!20!black,draw=mygreen!30!black,fill=mygreen!25]
\tikzstyle{node hidden}=[node,blue!20!black,draw=myblue!30!black,fill=myblue!20]
\tikzstyle{node convol}=[node,orange!20!black,draw=myorange!30!black,fill=myorange!20]
\tikzstyle{node out}=[node,red!20!black,draw=myred!30!black,fill=myred!20]
\tikzstyle{connect}=[thick,mydarkblue] 
\tikzstyle{connect arrow}=[-{Latex[length=4,width=3.5]},thick,mydarkblue,shorten <=0.5,shorten >=1]
\tikzset{ 
  node 1/.style={node in},
  node 2/.style={node hidden},
  node 3/.style={node out},
}

\usetikzlibrary{decorations.pathreplacing}
\usepackage{bbding} 
\usepackage{rotating}

\usepackage{media9}
\usepackage{multimedia}
\usepackage{animate}
\usepackage[frozencache=true,cachedir=minted-cache]{minted}
\newminted[python]{python}{frame=single}
\usepackage{fancyvrb}
\fvset{showspaces}

\usepackage{stmaryrd}

\usepackage[skins,theorems]{tcolorbox}
\tcbset{highlight math style={enhanced,
  colframe=blue,boxrule=0.4pt,colback=white,arc=0pt,boxrule=0.2pt}}

\usepackage{wrapfig}
\usepackage{capt-of}

\DeclareDocumentCommand{\bwfc}{ O{} O{} O{} O{l} O{20mm}}{
	\begin{wrapfigure}{#4}{#5}
		\begin{center}
			#1
			\captionof{figure}{#2}\label{fig:#3}
		\end{center}
\end{wrapfigure}
}

\DeclareDocumentCommand{\beqo}{ O{} O{} O{} }{
	\begin{equation}
	#1
	\label#3{eq:#2}
	\end{equation}
}
\DeclareDocumentCommand{\beqh}{ O{} O{} O{}  O{} }{
	\begin{equation}
	#1
	\hypertarget#3{#4}{\label#3{eq:#2}}
	\end{equation}
}

\DeclareDocumentCommand{\bsys}{ O{} O{} O{} }{
	\begin{empheq}[left=#1\empheqlbrace]{align}
	#2
	\label{eq:#3}
	\end{empheq}
}

\DeclareDocumentCommand{\bmat}{ O{} O{}  }{
	  \begin{bmatrix}
	  	\begin{array}{#2}
		#1
	\end{array}
	\end{bmatrix}
}
\DeclareDocumentCommand{\barr}{ O{} }{
	\begin{array}{c|c}
		#1
	\end{array}
}

\DeclareDocumentCommand{\NN}{O{}O{}}{\ensuremath{\mathcal{NN}_{#1}^{#2}}}
\DeclareDocumentCommand{\FNN}{O{}O{}O{}}{\ensuremath{\mathit{FNN}_{#2}^{#3}}}
\DeclareDocumentCommand{\MLP}{O{}O{}O{}}{\ensuremath{\mathit{MLP}_{#2}^{#3}}}
\DeclareDocumentCommand{\NLL}{O{}O{}O{}}{\ensuremath{\mathit{NLL}_{#1}^{#2}#3 }}

\DeclareDocumentCommand{\CNN}{O{}O{}O{}}{\ensuremath{\mathit{CNN}_{#2}^{#3}}}
\DeclareDocumentCommand{\RNN}{O{}O{}O{}}{\ensuremath{\mathit{RNN}_{#2}^{#3}}}
\DeclareDocumentCommand{\LSTM}{O{}O{}O{}}{\ensuremath{\mathit{LSTM}_{#2}^{#3}}}

\DeclareDocumentCommand{\KKNPP}{O{Nuclear Power Plant}}{Kashiwazaki-Kariwa #1}

\newcommand{\N}{\mathbb{N}}
\newcommand{\R}{\mathbb{R}}

\newcommand{\C}{\mathbb{C}}
\DeclareDocumentCommand{\opercal}{O{}O{}O{}}{\ensuremath{\mathcal{#1}_{#2}^{#3}}}
\DeclareDocumentCommand{\operbb}{O{}O{}O{}}{\ensuremath{\mathbb{#1}_{#2}^{#3}}}
\newcommand{\vct}[3]{\ensuremath{\boldsymbol{#1}_{#2}^{#3} }}
\DeclareDocumentCommand{\ebv}{ O{} O{} O{} }{\vct{e}{#1}{#2}#3}
\DeclareDocumentCommand{\ebbv}{ O{} O{} O{} }{\vct{\bar{e}}{#1}{#2}#3}
\DeclareDocumentCommand{\ebhv}{ O{} O{} O{} }{\vct{\hat{e}}{#1}{#2}#3}
\DeclareDocumentCommand{\debv}{ O{} O{} O{} }{\vct{\dot{e}}{#1}{#2}#3}
\DeclareDocumentCommand{\Phibv}{ O{} O{} O{} }{\vct{\Phi}{#1}{#2}#3}
\DeclareDocumentCommand{\ibv}{ O{} O{} O{} }{\vct{i}{#1}{#2}#3}

\DeclareDocumentCommand{\ione}{ O{} O{} }{\vct{i}{1}{#1}#2}
\DeclareDocumentCommand{\itwo}{ O{} O{} }{\vct{i}{2}{#1}#2}
\DeclareDocumentCommand{\ithree}{ O{} O{} }{\vct{i}{3}{#1}#2}
\DeclareDocumentCommand{\ivm}{ O{} O{} }{\vct{i}{m}{#1}#2}
\DeclareDocumentCommand{\ivn}{ O{} O{} }{\vct{i}{n}{#1}#2}
\DeclareDocumentCommand{\ivp}{ O{} O{} }{\vct{i}{p}{#1}#2}
\DeclareDocumentCommand{\ieta}{ O{} O{} }{\vct{i}{\eta}{#1}#2}
\DeclareDocumentCommand{\ix}{ O{} O{} }{\vct{i}{x}{#1}#2}
\DeclareDocumentCommand{\iy}{ O{} O{} }{\vct{i}{y}{#1}#2}

\DeclareDocumentCommand{\ir}{ O{} O{} O{} }{\vct{i}{r_{#1}}{#2}#3}
\DeclareDocumentCommand{\ith}{ O{} O{} O{} }{\vct{i}{\theta_{#1}}{#2}#3}
\DeclareDocumentCommand{\iz}{ O{} O{} O{} }{\vct{i}{z_{#1}}{#2}#3}
\DeclareDocumentCommand{\ixi}{ O{} }{\ibv[\xi]#1}
\DeclareDocumentCommand{\iXone}{ O{} }{\ibv[\chi_1]#1}
\DeclareDocumentCommand{\iXtwo}{ O{} }{\ibv[\chi_2]#1}
\DeclareDocumentCommand{\exi}{ O{} O{} }{\vct{e}{\xi}{#1}#2}
\DeclareDocumentCommand{\eXone}{ O{} }{\ebv[\chi_1]#1}
\DeclareDocumentCommand{\eXtwo}{ O{} }{\ebv[\chi_2]#1}
\DeclareDocumentCommand{\pG}{O{G}}{\pv[#1]}
\DeclareDocumentCommand{\pGp}{O{G} O{'}}{\pv[#1][#2]}
\DeclareDocumentCommand{\psigma}{O{\Sigma}}{\pv[#1]}
\DeclareDocumentCommand{\xG}{O{G}}{\xv[#1]}
\DeclareDocumentCommand{\xGo}{O{G_0}}{\xv[#1]}

\DeclareDocumentCommand{\xGp}{O{G} O{'}}{\xv[#1][#2]}
\DeclareDocumentCommand{\xGop}{O{G_0} O{'}}{\xv[#1][#2]}
\DeclareDocumentCommand{\xsigma}{O{\Sigma}}{\xv[#1]}
\DeclareDocumentCommand{\xsigmao}{O{\Sigma_0}}{\xv[#1]}
\DeclareDocumentCommand{\uG}{O{G}}{\uv[#1]}
\DeclareDocumentCommand{\uGo}{O{G_o}}{\uv[#1]}
\DeclareDocumentCommand{\uGp}{O{G}O{'}}{\uv[#1][#2]}
\DeclareDocumentCommand{\uGop}{O{G_0} O{'}}{\uv[#1][#2]}
\DeclareDocumentCommand{\uGopp}{O{G_0} O{''}}{\uv[#1][#2]}
\DeclareDocumentCommand{\uGe}{O{Ge} O{}}{u_{#1}#2}
\DeclareDocumentCommand{\uGep}{O{Ge} O{'}}{u_{#1}^{#2}}
\DeclareDocumentCommand{\uGepp}{O{Ge} O{''}}{u_{#1}^{#2}}

\DeclareDocumentCommand{\uGoe}{O{G_0e} O{}}{u_{#1}#2}
\DeclareDocumentCommand{\uGoep}{O{G_0e} O{'}}{u_{#1}^{#2}}
\DeclareDocumentCommand{\uGoepp}{O{G_0e} O{''}}{u_{#1}^{#2}}

\DeclareDocumentCommand{\usigma}{O{\Sigma}}{\uv[#1]}
\DeclareDocumentCommand{\uGsigma}{O{G\Sigma}}{\uv[#1]}
\DeclareDocumentCommand{\uGsigmap}{O{G\Sigma} O{'}}{\uv[#1][#2]}
\DeclareDocumentCommand{\uGsigmapp}{O{G\Sigma} O{''}}{\uv[#1][#2]}

\DeclareDocumentCommand{\usigmao}{O{\Sigma_0}}{\uv[#1]}
\DeclareDocumentCommand{\uGosigma}{O{G_0\Sigma_0}}{\uv[#1]}
\DeclareDocumentCommand{\uGosigmap}{O{G_0\Sigma_0} O{'}}{\uv[#1][#2]}
\DeclareDocumentCommand{\uGosigmapp}{O{G_0\Sigma_0} O{''}}{\uv[#1][#2]}

\DeclareDocumentCommand{\uGperp}{O{G\vdash}}{\uv[#1]}
\DeclareDocumentCommand{\uGperpp}{O{G\vdash} O{'}}{\uv[#1][#2]}
\DeclareDocumentCommand{\uGperppp}{O{G\vdash} O{''}}{\uv[#1][#2]}

\DeclareDocumentCommand{\thetam}{ O{} O{} O{} }{\tens{\bbtheta}{#1}{\wedge}#3}
\DeclareDocumentCommand{\thetamp}{ O{} O{} O{} }{\tens{\bbtheta}{#1}{\wedge'}#3}
\DeclareDocumentCommand{\thetav}{ O{} O{} O{} }{\vct{\theta}{#1}{#2}#3}
\DeclareDocumentCommand{\deltathetav}{ O{} O{} O{} }{\vct{\delta\theta}{#1}{#2}#3}
\DeclareDocumentCommand{\deltahatthetav}{ O{} O{} O{} }{\vct{\delta\hat{\theta}}{#1}{#2}#3}

\DeclareDocumentCommand{\deltav}{ O{} O{} O{} }{\vct{\delta}{#1}{#2}#3}
\DeclareDocumentCommand{\thetavp}{ O{} O{} }{\vct{\theta}{#1}{'}#2}
\DeclareDocumentCommand{\thetavd}{ O{} O{} }{\vct{\dot{\theta}}{#1}{#2}}
\DeclareDocumentCommand{\thetavdd}{ O{} O{} }{\vct{\ddot{\theta}}{#1}{#2}}
\DeclareDocumentCommand{\hatthetav}{ O{} O{} }{\vct{\hat{\theta}}{#1}{#2}}
\DeclareDocumentCommand{\hatomegav}{ O{} O{} }{\vct{\hat{\omega}}{#1}{#2}}
\DeclareDocumentCommand{\hatphiv}{ O{} O{} }{\vct{\hat{\phi}}{#1}{#2}}

\DeclareDocumentCommand{\barthetav}{ O{} O{} }{\vct{\bar{\theta}}{#1}{#2}}
\DeclareDocumentCommand{\thetavstar}{ O{} O{} }{\vct{\theta}{#1}{*}}
\DeclareDocumentCommand{\Thetav}{ O{} O{} }{\vct{\Theta}{#1}{#2}}
\DeclareDocumentCommand{\Thetavd}{ O{} O{} }{\vct{\dot{\Theta}}{#1}{#2}}
\DeclareDocumentCommand{\Thetavdd}{ O{} O{} }{\vct{\ddot{\Theta}}{#1}{#2}}
\DeclareDocumentCommand{\thsigmap}{ O{} }{\vct{\theta}{\Sigma}{'}#1}
\DeclareDocumentCommand{\thsigmapp}{ O{} }{\vct{\theta}{\Sigma}{''}#1}
\DeclareDocumentCommand{\thsigma}{ O{\Sigma} O{} O{} }{\vct{\theta}{#1}{#2}#3}
\DeclareDocumentCommand{\thetae}{ O{e} O{} }{\theta_{#1}#2}
\DeclareDocumentCommand{\thetaep}{ O{e} O{'} }{\theta_{#1}^{#2}}
\DeclareDocumentCommand{\thetavperp}{ O{\vdash} O{} }{\vct{\theta}{\vdash}{#2}}
\DeclareDocumentCommand{\thetavzeroperp}{ O{\vdash} O{} }{\vct{\theta}{0\vdash}{#2}}

\DeclareDocumentCommand{\dduGv}{ O{} O{} O{} }{\vct{\ddot{u}}{G}{#2}#3}
\DeclareDocumentCommand{\dduGsigma}{ O{} O{} O{} }{\vct{\ddot{u}}{G\Sigma}{#2}#3}
\DeclareDocumentCommand{\ddthetav}{ O{} O{} O{} }{\vct{\ddot{\theta}}{#1}{#2}#3}

\DeclareDocumentCommand{\pthp}{ O{} }{\ensuremath{\pdp{\theta_{#1}}} }
\DeclareDocumentCommand{\stressref}{O{0}}{\ensuremath{\stress \phantom{}_{\text{#1}}}}

\DeclareDocumentCommand{\hv}{ O{} O{} O{} }{\vct{h}{#1}{#2}#3}
\DeclareDocumentCommand{\hhv}{ O{} O{} O{} }{\vct{\hat{h}}{#1}{#2}#3}
\DeclareDocumentCommand{\ov}{ O{} O{} O{} }{\vct{o}{#1}{#2}#3}
\DeclareDocumentCommand{\yv}{ O{} O{} O{} }{\vct{y}{#1}{#2}#3}
\DeclareDocumentCommand{\yhv}{ O{} O{} O{} }{\vct{\hat{y}}{#1}{#2}#3}
\DeclareDocumentCommand{\ytv}{ O{} O{} O{} }{\vct{\tilde{y}}{#1}{#2}#3}
\DeclareDocumentCommand{\yvdot}{ O{} O{} O{} }{\vct{\dot{y}}{#1}{#2}#3}
\DeclareDocumentCommand{\dyv}{ O{} O{} O{} }{\vct{dy}{#1}{#2}#3}

\DeclareDocumentCommand{\ddyv}{ O{} O{} O{} }{\vct{\ddot{y}}{#1}{#2}#3}
\DeclareDocumentCommand{\pv}{ O{} O{} O{} }{\vct{p}{#1}{#2}#3}
\DeclareDocumentCommand{\dpv}{ O{} O{} O{} }{\vct{dp}{#1}{#2}#3}
\DeclareDocumentCommand{\pt}{ O{} O{} O{} }{\tens{p}{#1}{#2}#3}
\DeclareDocumentCommand{\Xv}{ O{} O{} O{} }{\vct{X}{#1}{#2}#3}
\DeclareDocumentCommand{\Xt}{ O{} O{} O{} }{\tens{\mathbb{X}}{#1}{#2}#3}
\DeclareDocumentCommand{\Xm}{ O{} O{} O{} }{\tens{\mathbf{X}}{#1}{#2}#3}
\DeclareDocumentCommand{\Yv}{ O{} O{} O{} }{\vct{Y}{#1}{#2}#3}
\DeclareDocumentCommand{\Ym}{ O{} O{} O{} }{\tens{\mathbf{Y}}{#1}{#2}#3}
\DeclareDocumentCommand{\Yt}{ O{} O{} O{} }{\tens{\mathbb{Y}}{#1}{#2}#3}

\DeclareDocumentCommand{\yt}{ O{} O{} O{} }{\tens{\mathbb{y}}{#1}{#2}#3}

\DeclareDocumentCommand{\Xhv}{ O{} O{} O{} }{\vct{\hat X}{#1}{#2}#3}
\DeclareDocumentCommand{\Zhv}{ O{} O{} O{} }{\vct{\hat Z}{#1}{#2}#3}
\DeclareDocumentCommand{\xv}{ O{} O{} O{} }{\vct{x}{#1}{#2}#3}
\DeclareDocumentCommand{\xhv}{ O{} O{} O{} }{\vct{\hat x}{#1}{#2}#3}
\DeclareDocumentCommand{\zhv}{ O{} O{} O{} }{\vct{\hat z}{#1}{#2}#3}
\DeclareDocumentCommand{\dxv}{ O{} O{} O{} }{\vct{dx}{#1}{#2}#3}
\DeclareDocumentCommand{\xvdot}{ O{} O{} O{} }{\vct{\dot{x}}{#1}{#2}#3}
\DeclareDocumentCommand{\Xtv}{ O{} O{} O{} }{\vct{\tilde{X}}{#1}{#2}#3}
\DeclareDocumentCommand{\xgv}{ O{} O{} O{} }{\vct{x}{g}{#1}#2}
\DeclareDocumentCommand{\ddxgv}{ O{} O{} O{} }{\vct{\ddot{x}}{g}{#1}#2}
\DeclareDocumentCommand{\tv}{ O{} O{} O{} }{\vct{t}{#1}{#2}#3}
\DeclareDocumentCommand{\fv}{ O{} O{} O{} }{\vct{f}{#1}{#2}#3}
\DeclareDocumentCommand{\fvdot}{ O{} O{} O{} }{\vct{\dot{f}}{#1}{#2}#3}
\DeclareDocumentCommand{\fvi}{ O{} O{} O{} }{\vct{f}{#1}{-1}#3}
\DeclareDocumentCommand{\ftv}{ O{} O{} O{} }{\vct{\tilde{f}}{#1}{#2}#3}
\DeclareDocumentCommand{\fhv}{ O{} O{} O{} }{\vct{\hat{f}}{#1}{#2}#3}
\DeclareDocumentCommand{\Et}{ O{} O{} O{} }{\tens{\mathbb{E}}{#1}{#2}#3}

\DeclareDocumentCommand{\Ft}{ O{} O{} O{} }{\tens{\mathbb{F}}{#1}{#2}#3}
\DeclareDocumentCommand{\Fm}{ O{} O{} O{} }{\tens{\mathbf{F}}{#1}{#2}#3}

\DeclareDocumentCommand{\kappat}{ O{} O{}}{\tens{\mathbb{k}}{#1}{#2}}
\DeclareDocumentCommand{\kv}{ O{} O{} O{} }{\vct{k}{#1}{#2}#3}
\DeclareDocumentCommand{\khv}{ O{} O{} O{} }{\vct{\hat{k}}{#1}{#2}#3}
\DeclareDocumentCommand{\sv}{ O{} O{} O{} }{\vct{s}{#1}{#2}#3}
\DeclareDocumentCommand{\shv}{ O{} O{} O{} }{\vct{\hat s}{#1}{#2}#3}
\DeclareDocumentCommand{\cv}{ O{} O{} O{} }{\vct{c}{#1}{#2}#3}
\DeclareDocumentCommand{\Cv}{ O{} O{} O{} }{\vct{C}{#1}{#2}#3}
\DeclareDocumentCommand{\Sv}{ O{} O{} O{} }{\vct{S}{#1}{#2}#3}
\DeclareDocumentCommand{\Nv}{ O{} O{} O{} }{\vct{N}{#1}{#2}#3}
\DeclareDocumentCommand{\muv}{ O{} O{} O{} }{\vct{\mu}{#1}{#2}#3}
\DeclareDocumentCommand{\hatmuv}{ O{} O{} O{} }{\vct{\hat\mu}{#1}{#2}#3}
\DeclareDocumentCommand{\mubv}{ O{} O{} O{} }{\vct{\bar{\mu}}{#1}{#2}#3}
\DeclareDocumentCommand{\dcv}{ O{} O{} O{} }{\vct{\dot{c}}{#1}{#2}#3}
\DeclareDocumentCommand{\ddcv}{ O{} O{} O{} }{\vct{\ddot{c}}{#1}{#2}#3}
\DeclareDocumentCommand{\ctv}{ O{} O{} O{} }{\vct{\tilde{c}}{#1}{#2}#3}
\DeclareDocumentCommand{\cbv}{ O{} O{} O{} }{\vct{\bar{c}}{#1}{#2}#3}
\DeclareDocumentCommand{\btv}{ O{} O{} O{} }{\vct{\tilde{b}}{#1}{#2}#3}
\DeclareDocumentCommand{\btt}{ O{} O{} O{} }{\tens{\tilde{\mathbbm{b}}}{#1}{#2}#3}
\DeclareDocumentCommand{\bbv}{ O{} O{} O{} }{\vct{\bar{b}}{#1}{#2}#3}
\DeclareDocumentCommand{\bbt}{ O{} O{} O{} }{\tens{\bar{\mathbbm{b}}}{#1}{#2}#3}
\DeclareDocumentCommand{\Sigmat}{ O{} O{} O{} }{\tens{\boldsymbol{\Sigma}}{#1}{#2}#3}
\DeclareDocumentCommand{\hatSigmat}{ O{} O{} O{} }{\tens{\boldsymbol{\hat\Sigma}}{#1}{#2}#3}
\DeclareDocumentCommand{\Ct}{ O{} O{} O{} }{\tens{\mathbb{C}}{#1}{#2}#3}
\DeclareDocumentCommand{\Cht}{ O{} O{} O{} }{\tens{\hat{\mathbb{C}}}{#1}{#2}#3}
\DeclareDocumentCommand{\dv}{ O{} O{} O{} }{\vct{d}{#1}{#2}#3}
\DeclareDocumentCommand{\Dt}{ O{} O{} O{} }{\tens{\mathbb{D}}{#1}{#2}#3}
\DeclareDocumentCommand{\DX}{ O{} }{\mathcal{D}_{\mathcal{X}}}
\DeclareDocumentCommand{\DXY}{ O{} }{\mathcal{D}_{\mathcal{XY}}}
\DeclareDocumentCommand{\Data}{ O{} }{\mathcal{D}}
\DeclareDocumentCommand{\DCY}{ O{} }{\mathcal{D}_{\mathcal{CY}}}
\DeclareDocumentCommand{\vV}{ O{} O{} O{} }{\vct{v}{#1}{#2}#3}
\DeclareDocumentCommand{\dvV}{ O{} O{} O{} }{\vct{dv}{#1}{#2}#3}
\DeclareDocumentCommand{\vt}{ O{} O{} O{} }{\tens{\mathbbm{v}}{#1}{#2}#3}
\DeclareDocumentCommand{\bv}{ O{} O{} O{} }{\vct{b}{#1}{#2}#3}
\DeclareDocumentCommand{\deltabv}{ O{} O{} O{} }{\vct{\delta b}{#1}{#2}#3}
\DeclareDocumentCommand{\bhv}{ O{} O{} O{} }{\vct{\hat b}{#1}{#2}#3}
\DeclareDocumentCommand{\bt}{ O{} O{} O{} }{\tens{\mathbbm{b}}{#1}{#2}#3}
\DeclareDocumentCommand{\Bm}{ O{} O{} O{} }{\tens{\mathbf{B}}{#1}{#2}#3}
\DeclareDocumentCommand{\Bt}{ O{} O{} O{} }{\tens{\mathbb{B}}{#1}{#2}#3}
\DeclareDocumentCommand{\St}{ O{} O{} O{} }{\tens{\mathbb{S}}{#1}{#2}#3}

\DeclareDocumentCommand{\Vv}{ O{} O{} O{} }{\vct{V}{#1}{#2}#3}
\DeclareDocumentCommand{\qv}{ O{} O{} O{} }{\vct{q}{#1}{#2}#3}
\DeclareDocumentCommand{\Qt}{ O{} O{} O{} }{\tens{\mathbb{Q}}{#1}{#2}#3}
\DeclareDocumentCommand{\Qm}{ O{} O{} O{} }{\tens{Q}{#1}{#2}#3}
\DeclareDocumentCommand{\Qtdot}{ O{} O{} O{} }{\tens{\dot{\mathbb{Q}}}{#1}{#2}#3}
\DeclareDocumentCommand{\qvF}{ O{} O{} O{} }{\vct{\hat{q}}{#1}{#2}#3}
\DeclareDocumentCommand{\ddqv}{ O{} O{} O{} }{\vct{\ddot{q}}{#1}{#2}#3}
\DeclareDocumentCommand{\dqv}{ O{} O{} O{} }{\vct{\dot{q}}{#1}{#2}#3}
\DeclareDocumentCommand{\deltaqv}{ O{} O{} O{} }{\vct{\delta q}{#1}{#2}#3}
\DeclareDocumentCommand{\av}{ O{} O{} O{} }{\vct{a}{#1}{#2}#3}
\DeclareDocumentCommand{\avf}{ O{} O{} O{} }{\vct{\tilde{a}}{#1}{#2}#3}
\DeclareDocumentCommand{\at}{ O{} O{} O{} }{\tens{\mathbbm{a}}{#1}{#2}#3}
\DeclareDocumentCommand{\Gt}{ O{} O{} O{} }{\tens{\mathbb{G}}{#1}{#2}#3}
\DeclareDocumentCommand{\Tt}{ O{} O{} O{} }{\tens{\mathbb{T}}{#1}{#2}#3}
\DeclareDocumentCommand{\Gt}{ O{} O{} O{} }{\tens{\mathbb{G}}{#1}{#2}#3}
\DeclareDocumentCommand{\Gttilde}{ O{} O{} O{} }{\tens{\mathbb{\tilde{G}}}{#1}{#2}#3}
\DeclareDocumentCommand{\Vttilde}{ O{} O{} O{} }{\tens{\mathbb{\tilde{V}}}{#1}{#2}#3}
\DeclareDocumentCommand{\Am}{ O{} O{} O{} }{\tens{A}{#1}{#2}#3}
\DeclareDocumentCommand{\At}{ O{} O{} O{} }{\tens{\mathbb{A}}{#1}{#2}#3}
\DeclareDocumentCommand{\Adt}{ O{} O{} O{} }{\tens{\dot{A}}{#1}{#2}#3}
\DeclareDocumentCommand{\Av}{ O{} O{} O{} }{\vct{A}{#1}{#2}#3}
\DeclareDocumentCommand{\wv}{ O{} O{} O{} }{\vct{w}{#1}{#2}#3}
\DeclareDocumentCommand{\whv}{ O{} O{} O{} }{\vct{\hat{w}}{#1}{#2}#3}
\DeclareDocumentCommand{\wbv}{ O{} O{} O{} }{\vct{\bar{w}}{#1}{#2}#3}
\DeclareDocumentCommand{\dwv}{ O{} O{} O{} }{\vct{\dot{w}}{#1}{#2}#3}
\DeclareDocumentCommand{\zv}{ O{} O{} O{} }{\vct{z}{#1}{#2}#3}
\DeclareDocumentCommand{\Zv}{ O{} O{} O{} }{\vct{Z}{#1}{#2}#3}
\DeclareDocumentCommand{\fm}{ O{} O{} O{} }{\vct{f}{#1}{#2}#3}
\DeclareDocumentCommand{\nv}{ O{} O{} O{} }{\vct{n}{#1}{#2}#3}
\DeclareDocumentCommand{\Nv}{ O{} O{} O{} }{\vct{N}{#1}{#2}#3}
\DeclareDocumentCommand{\Bv}{ O{} O{} O{} }{\vct{B}{#1}{#2}#3}
\DeclareDocumentCommand{\mv}{ O{} O{} O{} }{\vct{m}{#1}{#2}#3}
\DeclareDocumentCommand{\mhv}{ O{} O{} O{} }{\vct{\hat{m}}{#1}{#2}#3}
\DeclareDocumentCommand{\qv}{ O{} O{} O{} }{\vct{q}{#1}{#2}#3}
\DeclareDocumentCommand{\gv}{ O{} O{} O{} }{\vct{g}{#1}{#2}#3}
\DeclareDocumentCommand{\ghv}{ O{} O{} O{} }{\vct{\hat g}{#1}{#2}#3}
\DeclareDocumentCommand{\tauv}{ O{} O{} O{} }{\vct{\tau}{#1}{#2}#3}
\DeclareDocumentCommand{\tauvsigma}{ O{} O{} O{} }{\vct{\tau}{\Sigma}{#2}#3}
\DeclareDocumentCommand{\Nv}{ O{} O{} O{} }{\vct{N}{#1}{#2}#3}
\DeclareDocumentCommand{\Nm}{ O{} O{} O{} }{\tens{N}{#1}{#2}#3}
\DeclareDocumentCommand{\Rv}{ O{} O{} O{} }{\vct{R}{#1}{#2}#3}
\DeclareDocumentCommand{\Rm}{ O{} O{} O{} }{\tens{R}{#1}{#2}#3}
\DeclareDocumentCommand{\Rt}{ O{} O{} O{} }{\tens{\mathbb{R}}{#1}{#2}#3}
\DeclareDocumentCommand{\dRt}{ O{} O{} O{} }{\tens{\dot{\mathbb{R}}}{#1}{#2}#3}
\DeclareDocumentCommand{\Omegat}{ O{} O{} O{} }{\tens{\dOmega}{#1}{#2}#3}
\DeclareDocumentCommand{\OmegatT}{ O{} O{} O{} }{\tens{\dOmega}{#1}{T#2}#3}
\DeclareDocumentCommand{\omegav}{ O{} O{} O{} }{\vct{\omega}{#1}{#2}#3}
\DeclareDocumentCommand{\dOmegat}{ O{} O{} O{} }{\tens{\dot{\dOmega}}{#1}{#2}#3}

\DeclareDocumentCommand{\Mv}{ O{} O{} O{} }{\vct{M}{#1}{#2}#3}
\DeclareDocumentCommand{\lambdav}{ O{} O{} O{} }{\vct{\lambda}{#1}{#2}#3}
\DeclareDocumentCommand{\Lambdav}{ O{} O{} O{} }{\vct{\Lambda}{#1}{#2}#3}
\DeclareDocumentCommand{\Lambdat}{ O{} O{} O{} }{\tens{\Lambda}{#1}{#2}#3}
\DeclareDocumentCommand{\Fv}{ O{} O{} O{} }{\vct{F}{#1}{#2}#3}
\DeclareDocumentCommand{\Gv}{ O{} O{} O{} }{\vct{G}{#1}{#2}#3}
\DeclareDocumentCommand{\Qv}{ O{} O{} O{} }{\vct{Q}{#1}{#2}#3}
\DeclareDocumentCommand{\nvsigma}{ O{} }{\vct{n}{\Sigma}{}}
\DeclareDocumentCommand{\xiv}{ O{} O{} O{} }{\vct{xi}{#1}{#2}#3}
\DeclareDocumentCommand{\uv}{ O{} O{} O{} }{\vct{u}{#1}{#2}#3}
\DeclareDocumentCommand{\uhv}{ O{} O{} O{} }{\vct{\hat{u}}{#1}{#2}#3}
\DeclareDocumentCommand{\rv}{ O{} O{} O{} }{\vct{r}{#1}{#2}#3}
\DeclareDocumentCommand{\rhv}{ O{} O{} O{} }{\vct{\hat{r}}{#1}{#2}#3}
\DeclareDocumentCommand{\Uv}{ O{} O{} O{} }{\vct{U}{#1}{#2}#3}
\DeclareDocumentCommand{\Ut}{ O{} O{} O{} }{\tens{\mathbb{U}}{#1}{#2}#3}
\DeclareDocumentCommand{\Uvtilde}{ O{} O{} O{} }{\vct{\tilde{U}}{#1}{#2}#3}
\DeclareDocumentCommand{\dUt}{ O{} O{} O{} }{\tens{\dot{\mathbb{U}}}{#1}{#2}#3}
\DeclareDocumentCommand{\Um}{ O{} O{} O{} }{\tens{U}{#1}{#2}#3}
\DeclareDocumentCommand{\Vt}{ O{} O{} O{} }{\tens{\mathbb{V}}{#1}{#2}#3}
\DeclareDocumentCommand{\Vht}{ O{} O{} O{} }{\tens{\hat{\mathbb{V}}}{#1}{#2}#3}
\DeclareDocumentCommand{\Mt}{ O{} O{} O{} }{\tens{\mathbb{M}}{#1}{#2}#3}
\DeclareDocumentCommand{\Mvtilde}{ O{} O{} O{} }{\vct{\tilde{M}}{#1}{#2}#3}
\DeclareDocumentCommand{\unx}{ O{} }{\vct{\hat{u}}{n}{#1}\pdp{\xv}}
\DeclareDocumentCommand{\duv}{ O{} O{} O{} }{\vct{\dot{u}}{#1}{#2}#3}
\DeclareDocumentCommand{\dduv}{ O{} O{} O{} }{\vct{\ddot{u}}{#1}{#2}#3}
\DeclareDocumentCommand{\psiv}{ O{} O{} O{} }{\vct{\psi}{#1}{#2}#3}
\DeclareDocumentCommand{\psihv}{ O{} O{} O{} }{\vct{\hat\psi}{#1}{#2}#3}
\DeclareDocumentCommand{\phiv}{ O{} O{} O{} }{\vct{\phi}{#1}{#2}#3}
\DeclareDocumentCommand{\varphiv}{ O{} O{} O{} }{\vct{\varphi}{#1}{#2}#3}
\DeclareDocumentCommand{\Phiv}{ O{} O{} O{} }{\vct{\Phi}{#1}{#2}#3}
\DeclareDocumentCommand{\Phivt}{ O{} O{} O{} }{\vct{\tilde{\Phi}}{#1}{#2}#3}
\DeclareDocumentCommand{\dPsiv}{ O{} O{} O{} }{\vct{\dot{\Psi}}{#1}{#2}#3}
\DeclareDocumentCommand{\Psiv}{ O{} O{} O{} }{\vct{\Psi}{#1}{#2}#3}
\DeclareDocumentCommand{\dPhiv}{ O{} O{} O{} }{\vct{\dot{\Phi}}{#1}{#2}#3}
\DeclareDocumentCommand{\phim}{ O{} O{} O{} }{\tens{\Phi}{#1}{#2}#3}
\DeclareDocumentCommand{\vphiv}{ O{} O{} O{} }{\vct{\varphi}{#1}{#2}#3}

\DeclareDocumentCommand{\XmYC}{ O{x} O{} O{x} O{O}}{
	\pdp{\tens{\mathbbm{#1}}{#2}{}-\tens{\mathbbm{#3}}{#4}{}}^{\wedge}
}
\DeclareDocumentCommand{\XmC}{O{ }}{\tens{\mathbbm{x}}{#1}{\wedge}}

\DeclareDocumentCommand{\imC}{O{m}}{\tens{\mathbbm{i}}{#1}{\wedge}}

\newcommand{\zerov}{\vct{0}{}{}}

\newcommand{\pdp}[1]{\ensuremath{\left(#1\right)}}
\newcommand{\psp}[1]{\ensuremath{\left[#1\right]}}

\newcommand{\pbp}[1]{\ensuremath{\left\lbrace #1\right\rbrace }}

\newcommand{\davvr}[6]{\ensuremath{\left(\vct{#1}{#2}{#3},\vct{#4}{#5}{#6}\right)}}
\newcommand{\davsr}[6]{\ensuremath{\left(\vct{#1}{#2}{#3};#4_{#5}^{#6}\right)}}

\DeclareDocumentCommand{\xiyi}{O{x} O{y} O{i} O{i}}{\davvr{#1}{#3}{}{#2}{#4}{}}
\DeclareDocumentCommand{\xifxi}{O{x} O{f} O{i} O{i}}{\pdp{\vct{#1}{#3}{},\vct{#2}{}{}\pdp{\vct{#1}{#4}{}} }}

\newcommand{\pxtp}{\ensuremath{\davsr{x}{}{}{t}{}{}}}
\newcommand{\pXtp}{\ensuremath{\davsr{X}{}{}{t}{}{}}}

\newcommand{\ptp}{\ensuremath{\pdp{t}}}

\DeclareDocumentCommand{\pvp}{O{v} O{} O{}}{\ensuremath{\pdp{\vct{#1}{#2}{#3}}}}
\DeclareDocumentCommand{\pvxtp}{O{v} O{} O{}}{\ensuremath{\pdp{\vct{#1}{#2}{#3}\pxtp}}}
\DeclareDocumentCommand{\pvXtp}{O{v} O{} O{}}{\ensuremath{\pdp{\vct{#1}{#2}{#3}\pXtp}}}

\DeclareDocumentCommand{\Yv}{ O{} O{} O{} }{\vct{Y}{#1}{#2}#3}

\DeclareDocumentCommand{\otp}{ O{ } O{ } }{\ensuremath{#1\otimes #2}}
\DeclareDocumentCommand{\sotp}{ O{ } O{ } }{\ensuremath{#1\otimes_s#2}}
\DeclareDocumentCommand{\aotp}{ O{ } O{ } }{\ensuremath{#1\otimes_a#2}}

\newcommand{\tens}[3]{\ensuremath{\boldsymbol{#1}\phantom{}_{#2}^{#3}}}

\newcommand{\tensdot}[3]{\ensuremath{\boldsymbol{\dot{#1}}_{#2}^{#3}}}

\DeclareDocumentCommand{\Jin}{ O{} O{} }{\ensuremath{\tens{\mathbb{J}_{#1}}{}{#2}}}

\DeclareDocumentCommand{\Jinm}{ O{\rho} }{
	\ensuremath{\tens{\mathbb{J}_{}}{}{}\phantom{}_{#1}}
}
\DeclareDocumentCommand{\stress}{ O{} O{} }{\ensuremath{\tens{\dsigma}{#1}{#2}}}
\DeclareDocumentCommand{\stressdot}{ O{} O{} }{\ensuremath{\tensdot{\dsigma}{#1}{#2}}}
\DeclareDocumentCommand{\dstress}{ O{} O{} }{\ensuremath{\tens{\delta\dsigma}{#1}{#2}}}

\DeclareDocumentCommand{\strain}{ O{} O{}}{\ensuremath{\tens{\varepsilon}{#1}{#2}}}
\DeclareDocumentCommand{\straindot}{ O{} O{}}{\ensuremath{\tens{\dot{\varepsilon}}{#1}{#2}}}
\DeclareDocumentCommand{\deltastrain}{ O{} O{}}{\ensuremath{\tens{\delta\varepsilon}{#1}{#2}}}
\DeclareDocumentCommand{\dstrain}{ O{} O{}}{\ensuremath{d\tens{\varepsilon}{#1}{#2}}}
\DeclareDocumentCommand{\strainel}{O{}}{\ensuremath{\strain[#1][\text{el}]}}

\DeclareDocumentCommand{\deviator}{ O{} O{} }{\tens{\mathbbm{s}}{#1}{\dsigma#2}}

\DeclareDocumentCommand{\edev}{O{}}{\ensuremath{\tens{\mathbbm{e}}{#1}{}}}
\DeclareDocumentCommand{\epl}{O{}}{\ensuremath{\tens{\mathbbm{e}}{#1}{pl}}}
\DeclareDocumentCommand{\eel}{O{}}{\ensuremath{\tens{\mathbbm{e}}{#1}{el}}}
\DeclareDocumentCommand{\evol}{O{}}{\ensuremath{\varepsilon_{\text{vol}}^{#1}}}
\DeclareDocumentCommand{\devol}{O{}}{\ensuremath{\delta\varepsilon_{\text{vol}}^{#1}}}
\DeclareDocumentCommand{\evoldot}{O{}}{\ensuremath{\dot{\varepsilon}_{\text{vol}}^{#1}}}

\newcommand{\evolpl}{\ensuremath{\evol[\text{pl}]}}
\DeclareDocumentCommand{\devolel}{O{}}{\ensuremath{\delta\varepsilon_{\text{vol}}^{\text{el}}}}
\DeclareDocumentCommand{\evoldotel}{O{}}{\ensuremath{\dot{\varepsilon}_{\text{vol}}^{\text{el}}}}

\DeclareDocumentCommand{\evolpl}{O{}}{\ensuremath{\varepsilon_{\text{vol}}^{\text{pl}}}}
\DeclareDocumentCommand{\devolpl}{O{}}{\ensuremath{\delta\varepsilon_{\text{vol}}^{\text{pl}}}}
\DeclareDocumentCommand{\evoldotpl}{O{}}{\ensuremath{\dot{\varepsilon}_{\text{vol}}^{\text{pl}}}}
\DeclareDocumentCommand{\ebar}{O{} O{}}{\ensuremath{\bar{e}{#1}^{#2}}}
\DeclareDocumentCommand{\edev}{O{}}{\ensuremath{\varepsilon_\text{d}^{#1}}}
\DeclareDocumentCommand{\dedev}{O{}}{\ensuremath{\delta\varepsilon_d^{#1}}}
\newcommand{\dedevel}{\ensuremath{\delta\varepsilon_\text{d}^{\text{el}}}}
\newcommand{\dedevpl}{\ensuremath{\delta\varepsilon_\text{d}^{\text{pl}}}}
\DeclareDocumentCommand{\edvdot}{O{}}{\ensuremath{\dot{\varepsilon}_d^{#1}}}
\DeclareDocumentCommand{\edvpldot}{O{}}{\ensuremath{\edvdot[\text{pl}]}}
\DeclareDocumentCommand{\edveldot}{O{}}{\ensuremath{\edvdot[\text{el}]}}

\DeclareDocumentCommand{\edevdot}{O{}}{\ensuremath{\tensdot{\mathbbm{e}}{#1}{\text{el}}}}
\DeclareDocumentCommand{\edevpldot}{O{}}{\ensuremath{\tensdot{\mathbbm{e}}{#1}{\text{pl}}}}
\DeclareDocumentCommand{\edeveldot}{O{}}{\ensuremath{\tensdot{\mathbbm{e}}{#1}{\text{el}}}}

\DeclareDocumentCommand{\dedev}{O{}}{\ensuremath{\tens{\delta \mathbbm{e}}{#1}{}}}
\DeclareDocumentCommand{\dedevpl}{O{}}{\ensuremath{\tens{\delta \mathbbm{e}}{#1}{\text{pl}}}}
\DeclareDocumentCommand{\dedevel}{O{}}{\ensuremath{\tens{\delta \mathbbm{e}}{#1}{\text{el}}}}

\DeclareDocumentCommand{\fvm}{O{}}{\ensuremath{\sqrt{3J_{2}\pdp{\deviator-\tens{X}{}{}#1}}}}

\DeclareDocumentCommand{\backstress}{O {} O {} O{}}{\ensuremath{\tens{\mathbb{X}}{}{}}}
\DeclareDocumentCommand{\kbackstress}{O {} O {} O{}}{\ensuremath{\tens{\dalpha}{}{}}}
\DeclareDocumentCommand{\dbackstress}{O {} O {} O{}}{\ensuremath{\tens{\delta \mathbb{X}}{}{}}}
\DeclareDocumentCommand{\backstressdot}{O {} O {} O{}}{\ensuremath{\tensdot{\mathbb{X}}{}{}}}
\DeclareDocumentCommand{\dkbackstress}{O {} O {} O{}}{\ensuremath{\tens{\delta\dalpha}{}{}}}
\DeclareDocumentCommand{\backstressdot}{O {} O {} O{}}{\ensuremath{\tensdot{\mathbb{X}}{}{}}}
\DeclareDocumentCommand{\kbackstressdot}{O {} O {} O{}}{\ensuremath{\tensdot{\dalpha}{}{}}}

\newcommand{\norm}[1]{\ensuremath{\Vert #1 \Vert}}

\newcommand{\ID}{\ensuremath{\tens{\mathbb{I}}{}{}}}
\DeclareDocumentCommand{\IF}{O {} O {}}{\ensuremath{\tens{\mathbb{I}}{#1}{F}#2}}
\DeclareDocumentCommand{\IhF}{O {} O {}}{\ensuremath{\tens{\mathbb{\hat I}}{#1}{F}#2}}

\DeclareDocumentCommand{\RotQ}{ O{} O{} O{} O{}}{ \ensuremath{\tens{\mathbb{Q}}{}{}\davsr{#1}{#2}{}{#3}{}{}}
}
\DeclareDocumentCommand{\Pt}{ O{} O{} }{\ensuremath{\tens{\mathbb{P}}{#1}{#2}}}
\DeclareDocumentCommand{\Pm}{ O{} O{} }{\ensuremath{\tens{P}{#1}{#2}}}
\DeclareDocumentCommand{\Phm}{ O{} O{} }{\ensuremath{\tens{\hat{P}}{#1}{#2}}}
\DeclareDocumentCommand{\Pbm}{ O{} O{} }{\ensuremath{\tens{\bar{P}}{#1}{#2}}}

\DeclareDocumentCommand{\Dsign}{O{} O{}}{\tens{\Delta \dsigma}{#1}{#2}}
\DeclareDocumentCommand{\Dchihn}{O{} O{}}{\hidd{\Delta \dchi}{#1}{#2}}

\newcommand{\hidd}[3]{\ensuremath{\underset{\sim}{\smash{\boldsymbol{#1}}}\phantom{}_{#2}^{#3}}}

\DeclareDocumentCommand{\epsv}{ O{} O{} }{\ensuremath{\vct{\varepsilon}{#1}{#2}}}
\DeclareDocumentCommand{\etav}{ O{} O{} }{\ensuremath{\vct{\eta}{#1}{#2}}}

\DeclareDocumentCommand{\etah}{ O{} O{} }{\ensuremath{\hidd{\eta}{#1}{#2}}}
\DeclareDocumentCommand{\detah}{ O{} O{} }{\ensuremath{\hidd{\delta\eta}{#1}{#2}}}
\DeclareDocumentCommand{\etahdot}{ O{} O{} }{\ensuremath{\hidd{\dot{\eta}}{#1}{#2}}}

\DeclareDocumentCommand{\chih}{ O{} O{} }{\ensuremath{\hidd{\dchi}{#1}{#2}}}
\DeclareDocumentCommand{\chihdot}{ O{} O{} }{\ensuremath{\hidd{\dot{\dchi}}{#1}{#2}}}
\DeclareDocumentCommand{\dchih}{ O{} O{} }{\ensuremath{\hidd{\delta\dchi}{#1}{#2}}}

\DeclareDocumentCommand{\nablav}{O{x} O{}}{\ensuremath{\vct{\nabla}{#1}{#2}}}

\DeclareDocumentCommand{\dive}{ O{x} }{\ensuremath{\vct{\nabla}{#1}{}.}}
\DeclareDocumentCommand{\divergence}{ O{x} }{\ensuremath{\text{div}_{#1}}}

\DeclareDocumentCommand{\laplv}{ O{x} }{\ensuremath{\boldsymbol{\Delta}_{#1} }}

\DeclareDocumentCommand{\matd}{O{} O{t}}{\ensuremath{\frac{d#1}{d#2}}}
\DeclareDocumentCommand{\Depsn}{O{} O{}}{\tens{\Delta \varepsilon}{#1}{#2}}
\DeclareDocumentCommand{\Depsnel}{O{} O{}}{\tens{\Delta \varepsilon}{#1}{el}}
\DeclareDocumentCommand{\Depsnpl}{O{} O{}}{\tens{\Delta \varepsilon}{#1}{pl}}

\DeclareDocumentCommand{\pard}{O{} O{t}}{\ensuremath{\frac{\partial #1}{\partial #2}}}
\DeclareDocumentCommand{\pardd}{O{} O{t}}{\ensuremath{\frac{\partial #1\phantom{}^2}{\partial^2 #2}}}

\DeclareDocumentCommand{\partdt}{ O{} O{x} }{\ensuremath{\frac{\partial #1}{\partial t}} }

\DeclareDocumentCommand{\gammav}{O{} O{}}{\vct{\gamma}{#1}{#2}}

\DeclareDocumentCommand{\Gammav}{O{} O{}}{\vct{\Gamma}{#1}{#2}}

\newcommand{\E}[2]{\ensuremath{\mathbb{E}_{#1}\psp{#2}}}

\DeclareDocumentCommand{\iO}{O{t}}{
\ensuremath{\int_{\Omega_{#1}}}}

\DeclareDocumentCommand{\xtv}{ O{} O{} O{} }{\vct{\tilde{x}}{#1}{#2}#3}
\DeclareDocumentCommand{\xbv}{ O{} O{} O{} }{\vct{\bar{x}}{#1}{#2}#3}
\DeclareDocumentCommand{\ztv}{ O{} O{} O{} }{\vct{\tilde{z}}{#1}{#2}#3}
\DeclareDocumentCommand{\Ztv}{ O{} O{} O{} }{\vct{\tilde{Z}}{#1}{#2}#3}
\DeclareDocumentCommand{\zbv}{ O{} O{} O{} }{\vct{\bar{z}}{#1}{#2}#3}
\DeclareDocumentCommand{\Zbv}{ O{} O{} O{} }{\vct{\bar{Z}}{#1}{#2}#3}

\DeclareDocumentCommand{\xiv}{ O{} O{} O{} }{\vct{\xi}{#1}{#2}#3}
\DeclareDocumentCommand{\xitv}{ O{} O{} O{} }{\vct{\tilde{\xi}}{#1}{#2}#3}
\DeclareDocumentCommand{\xihv}{ O{} O{} O{} }{\vct{\hat{\xi}}{#1}{#2}#3}
\DeclareDocumentCommand{\xibv}{ O{} O{} O{} }{\vct{\bar{\xi}}{#1}{#2}#3}

\DeclareDocumentCommand{\pxv}{ O{} O{} O{}}{\pdp{\xv[#1][#2][#3]}}
\DeclareDocumentCommand{\pxtv}{ O{} O{} O{}}{\pdp{\xtv[#1][#2][#3]}}
\DeclareDocumentCommand{\pxtuv}{ O{} O{} O{}}{\pdp{\xtv[#1][#2][#3],u\pxtv[#1][#2][#3]}}
\DeclareDocumentCommand{\pxiv}{ O{} O{} O{}}{\pdp{\xiv[#1][#2][#3]}}
\DeclareDocumentCommand{\pxitv}{ O{} O{} O{}}{\pdp{\xitv[#1][#2][#3]}}
\DeclareDocumentCommand{\pxituv}{ O{} O{} O{}}{\pdp{\xitv[#1][#2][#3],u\pxitv[#1][#2][#3]}}
\DeclareDocumentCommand{\afc}{ O{} O{} O{}}{\vct{a}{}{}\pxtuv[#1][#2][#3]}
\DeclareDocumentCommand{\aftc}{ O{} O{} O{}}{\vct{\tilde{a}}{}{}\pxtuv[#1][#2][#3]}

\DeclareDocumentCommand{\Sset}{O{}}{\ensuremath{\mathcal{S}#1}}

\DeclareDocumentCommand{\Pr}{O{}}{\ensuremath{\mathit{P}#1}}
\DeclareDocumentCommand{\Sset}{O{} O{}}{\ensuremath{\mathcal{S}_{#1}^{#2}}}
\DeclareDocumentCommand{\Dset}{O{} O{}}{\ensuremath{\mathcal{D}_{#1}^{#2}}}
\DeclareDocumentCommand{\emean}{O{}}{\ensuremath{\mathbb{E}_{{\scriptscriptstyle{#1}}}}}
\DeclareDocumentCommand{\var}{O{}}{\ensuremath{\mathbb{V}_{{\scriptscriptstyle{#1}}}}}

\DeclareDocumentCommand{\hatemean}{O{}}{\ensuremath{\hat{\mathbb{E}}_{{\scriptscriptstyle{#1}}}}}
\DeclareDocumentCommand{\pxy}{O{}}{\ensuremath{p_{{\scriptscriptstyle{(\Xv,\Yv)}}}#1}}
\DeclareDocumentCommand{\px}{O{}}{\ensuremath{p_{{\scriptscriptstyle{\Xv}}}#1}}
\DeclareDocumentCommand{\qx}{O{}}{\ensuremath{q_{{\scriptscriptstyle{\Xv}}}#1}}
\DeclareDocumentCommand{\qy}{O{}}{\ensuremath{q_{{\scriptscriptstyle{\Yv}}}#1}}
\DeclareDocumentCommand{\qxy}{O{}}{\ensuremath{q_{{\scriptscriptstyle{(\Xv,\Yv)}}}#1}}
\DeclareDocumentCommand{\Pxy}{O{}}{\ensuremath{P_{{\scriptscriptstyle{(\Xv,\Yv)}}}#1}}
\DeclareDocumentCommand{\Px}{O{}}{\ensuremath{P_{{\scriptscriptstyle{\Xv}}}#1}}
\DeclareDocumentCommand{\hPxy}{O{}}{\ensuremath{\hat{P}_{{\scriptscriptstyle{(\Xv,\Yv)}}}#1}}
\DeclareDocumentCommand{\hpxy}{O{}}{\ensuremath{\hat{p}_{{\scriptscriptstyle{(\Xv,\Yv)}}}#1}}
\DeclareDocumentCommand{\hPx}{O{}}{\ensuremath{\hat{P}_{\Xv}#1}}

\DeclareDocumentCommand{\pxz}{O{}}{\ensuremath{p_{{\scriptscriptstyle{XZ}}}#1}}
\DeclareDocumentCommand{\qz}{O{}}{\ensuremath{q_{{\scriptscriptstyle{Z}}}#1}}
\DeclareDocumentCommand{\qxz}{O{}}{\ensuremath{q_{{\scriptscriptstyle{XZ}}}#1}}

\DeclareDocumentCommand{\XZv}{O{}}{\ensuremath{(\Xv,\Zv)}}

\DeclareDocumentCommand{\pxcz}{O{}}{\ensuremath{p_{{\scriptscriptstyle{X\vert Z}}}#1}}

\DeclareDocumentCommand{\pzcx}{O{}}{\ensuremath{p_{{\scriptscriptstyle{Z\vert X}}}#1}}
\DeclareDocumentCommand{\qzcx}{O{}}{\ensuremath{q_{{\scriptscriptstyle{Z\vert X}}}#1}}
\DeclareDocumentCommand{\qxcz}{O{}}{\ensuremath{q_{{\scriptscriptstyle{X\vert Z}}}#1}}

\DeclareDocumentCommand{\pux}{O{}}{\ensuremath{u_{{\scriptscriptstyle{X}}}#1}}
\DeclareDocumentCommand{\puy}{O{}}{\ensuremath{u_{{\scriptscriptstyle{Y}}}#1}}
\DeclareDocumentCommand{\puxy}{O{}}{\ensuremath{u_{{\scriptscriptstyle{XY}}}#1}}

\DeclareDocumentCommand{\pz}{O{}}{\ensuremath{p_{{\scriptscriptstyle{Z}}}#1}}
\DeclareDocumentCommand{\py}{O{}}{\ensuremath{p_{{\scriptscriptstyle{Y}}}#1}}
\DeclareDocumentCommand{\pycx}{O{}}{\ensuremath{p_{{\scriptscriptstyle{Y\vert X}}}#1}}
\DeclareDocumentCommand{\Pycx}{O{}}{\ensuremath{P_{{\scriptscriptstyle{\Yv\vert \Xv}}}#1}}

\DeclareDocumentCommand{\pxcy}{O{}}{\ensuremath{p_{{\scriptscriptstyle{X\vert Y}}}#1}}
\DeclareDocumentCommand{\qycx}{O{}}{\ensuremath{q_{{\scriptscriptstyle{Y\vert X}}}#1}}
\DeclareDocumentCommand{\puycx}{O{}}{\ensuremath{u_{{\scriptscriptstyle{Y\vert X}}}#1}}
\DeclareDocumentCommand{\xsimpd}{O{}}{\ensuremath{\vct{x}{}{}\sim\pdata[]}}
\DeclareDocumentCommand{\Wv}{O{} O{}}{\ensuremath{\vct{W}{#1}{#2}}}
\DeclareDocumentCommand{\DKL}{O{p} O{q}}{\ensuremath{\mathbb{D}_{KL}\psp{#1 : #2}}}

\newcommand{\Loss}{\ensuremath{\mathcal{L}}}

\DeclareDocumentCommand{\amin}{O{}}{\ensuremath{\underset{#1}{\text{arg min }}}}
\DeclareDocumentCommand{\amin}{O{}}{\ensuremath{\underset{#1}{\text{arg max }}}}

\DeclareDocumentCommand{\Sd}{O{} O{}}{\ensuremath{\mathbb{S}\pdp{#1\vert\vert#2}}}

\DeclareDocumentCommand{\DJS}{O{p} O{q}}{\ensuremath{\mathbb{D}_{JS}\pdp{#1\vert\vert #2}}}
\DeclareDocumentCommand{\kappav}{ O{} O{}}{\vct{\kappa}{#1}{#2}}
\DeclareDocumentCommand{\kappavdot}{ O{} O{}}{\vct{\dot{\kappa}}{#1}{#2}}
\DeclareDocumentCommand{\kappat}{ O{} O{}}{\tens{\mathbb{k}}{#1}{#2}}
\DeclareDocumentCommand{\Km}{ O{} O{}}{\tens{K}{#1}{#2}}
\DeclareDocumentCommand{\Kt}{ O{} O{}}{\tens{\mathbb{K}}{#1}{#2}}
\DeclareDocumentCommand{\Lt}{ O{} O{}}{\tens{\mathbb{L}}{#1}{#2}}
\DeclareDocumentCommand{\Lv}{ O{} O{}}{\vct{L}{#1}{#2}}
\DeclareDocumentCommand{\lT}{ O{} O{}}{\tens{\mathbbm{l}}{#1}{#2}}

\DeclareDocumentCommand{\Mm}{ O{} O{}}{\tens{M}{#1}{#2}}
\DeclareDocumentCommand{\Cm}{ O{} O{}}{\tens{\mathbb{C}}{#1}{#2}}
\DeclareDocumentCommand{\Wm}{ O{} O{}}{\tens{W}{#1}{#2}}
\DeclareDocumentCommand{\deltaWm}{ O{} O{}}{\tens{\delta W}{#1}{#2}}
\DeclareDocumentCommand{\Whm}{ O{} O{}}{\tens{\hat W}{#1}{#2}}
\DeclareDocumentCommand{\Rm}{ O{} O{}}{\tens{R}{#1}{#2}}
\DeclareDocumentCommand{\Sm}{ O{} O{}}{\tens{S}{#1}{#2}}
\DeclareDocumentCommand{\Hm}{ O{} O{}}{\tens{H}{#1}{#2}}
\DeclareDocumentCommand{\Ht}{ O{} O{}}{\tens{\mathbb{H}}{#1}{#2}}
\DeclareDocumentCommand{\Hhm}{ O{} O{}}{\tens{\hat{H}}{#1}{#2}}
\DeclareDocumentCommand{\Um}{ O{} O{}}{\tens{U}{#1}{#2}}
\DeclareDocumentCommand{\Jm}{ O{} O{}}{\tens{J}{#1}{#2}}
\DeclareDocumentCommand{\Jt}{ O{} O{}}{\tens{\mathbb{J}}{#1}{#2}}
\DeclareDocumentCommand{\FextF}{ O{} O{}}{\vct{\hat{F}}{#1}{#2}}
\DeclareDocumentCommand{\Fint}{ O{} O{\text{int}}}{\vct{F}{#1}{#2}}
\DeclareDocumentCommand{\Fext}{ O{} O{\text{ext}}}{\vct{F}{#1}{#2}}
\DeclareDocumentCommand{\diagKm}{ O{} O{}}{\mathsf{diag}\pdp{\Km[#1][#2]}}
\DeclareDocumentCommand{\diagMm}{ O{} O{}}{\mathsf{diag}\pdp{\Mm[#1][#2]}}
\DeclareDocumentCommand{\diagsc}{ O{}}{\mathsf{diag}\pdp{#1}}
\DeclareDocumentCommand{\rhod}{O{}}{\ensuremath{\rho_{#1}}}
\DeclareDocumentCommand{\rhoref}{O{}}{\ensuremath{\rho^{\text{ref}}_{#1}}}

\DeclareDocumentCommand{\Lla}{ O{\pdp{\xv}}}{\ensuremath{\lambda#1}}
\DeclareDocumentCommand{\Lmu}{ O{\pdp{\xv}}}{\ensuremath{\mu#1}}
\DeclareDocumentCommand{\diagsc}{ O{}}{\mathsf{diag}\pdp{#1}}

\usepackage{thmtools}

\DeclareDocumentCommand{\Hessian}{ O{} O{}}{\tens{\mathbb{H}}{#1}{#2}}

\DeclareDocumentCommand{\Estrain}{O {} O{}}{\ensuremath{\tens{\mathbb{E}}{#1}{#2}}}

\DeclareDocumentCommand{\Sstress}{O {} O{}}{\ensuremath{\tens{\mathbb{S}}{#1}{#2}}}

\DeclareDocumentCommand{\gradetah}{O{}}{\ensuremath{\tens{D}{\eta}{}#1}}

\DeclareDocumentCommand{\Hent}{O{}}{\ensuremath{\mathcal{H}_{#1}}}
\DeclareDocumentCommand{\hessian}{O{} O{}}{\ensuremath{\tens{\mathbb{H}}{#1}{#2}}}
\DeclareDocumentCommand{\MBL}{O{} O{} O{}}{\ensuremath{\mathcal{M}^{#3}\pdp{#1,#2}}}
\DeclareDocumentCommand{\KBL}{O{} O{}}{\ensuremath{\mathcal{K}\pdp{#1,#2}}}
\DeclareDocumentCommand{\FL}{O{}}{\ensuremath{\mathcal{F}\pdp{#1}}}

\newcommand{\be}{\begin{equation}}
\newcommand{\ee}{\end{equation}}
\newcommand{\bi}{\begin{itemize}}
\newcommand{\ei}{\end{itemize}}
\newcommand{\ben}{\begin{enumerate}}
\newcommand{\een}{\end{enumerate}}
\newcommand{\bea}{\begin{eqnarray}}
\newcommand{\eea}{\end{eqnarray}}

\usepackage{listings}
\usepackage{color}
\definecolor{dkgreen}{rgb}{0,0.6,0}
\definecolor{gray}{rgb}{0.5,0.5,0.5}
\definecolor{lgray}{rgb}{0.75,0.75,0.75}
\definecolor{mauve}{rgb}{0.58,0,0.82}
\definecolor{deepblue}{rgb}{0,0,0.5}
\definecolor{deepred}{rgb}{0.6,0,0}
\definecolor{deepgreen}{rgb}{0,0.5,0}
\definecolor{olive}{rgb}{0.3, 0.4, .1}
\definecolor{fore}{RGB}{249,242,215}
\definecolor{back}{RGB}{51,51,51}
\definecolor{title}{RGB}{255,0,90}
\definecolor{dgreen}{rgb}{0.,0.6,0.}
\definecolor{fgreen}{RGB}{0 212 85}
\definecolor{gold}{rgb}{1.,0.84,0.}
\definecolor{junglegreen}{cmyk}{0.99,0,0.52,0}
\definecolor{bluegreen}{cmyk}{0.85,0,0.33,0}
\definecolor{rawsienna}{cmyk}{0,0.72,1,0.45}
\definecolor{magenta}{cmyk}{0,1,0,0}
\definecolor{POLIMIgray}{RGB}{114,143,165}
\definecolor{IntenseBlue}{rgb}{0,0.7109,1.0000}
\definecolor{IntenseOrange}{rgb}{1.0000,0.4141,0.1875}
\definecolor{IntenseGreen}{rgb}{0.1211,0.7109,0.1445}
\definecolor{sinapsGreen}{RGB}{130,202,63}
\definecolor{sinapsDarkGreen}{RGB}{0,128,0}
\definecolor{sinapsLightGreen}{RGB}{208,237,179}

\usepackage{verbatim}
\usepackage{mathtools} 
\usepackage{newtxtext,newtxmath}
\usepackage{booktabs}

\newsavebox{\twosubbox}
\usepackage[numbers]{natbib}

\usepackage{lineno}
\usepackage{url}

\title{Physics-based super-resolved simulation of 3D elastic wave propagation adopting scalable Diffusion Transformer}

\author{
    Hugo Gabrielidis\\
        Université Paris-Saclay, CentraleSupélec, CNRS, ENS \\
        Laboratoire de Mécanique Paris-Saclay UMR 9026\\
        8-10 rue Joliot Curie, 91190 Gif-sur-Yvette, France\\
        Laboratoire Interdisciplinaire des Sciences du Numérique\\ CentraleSupélec, CNRS, Université Paris-Saclay\\
        rue du Belvédère, 91405 Orsay, France\\
        \texttt{hugo.gabrielidis@centralesupelec.fr} \\
        \And 
        Filippo Gatti\\
        Université Paris-Saclay, CentraleSupélec, CNRS, ENS \\
        Laboratoire de Mécanique Paris-Saclay UMR 9026\\
        8-10 rue Joliot Curie, 91190 Gif-sur-Yvette, France\\
        \texttt{filippo.gatti@centralesupelec.fr} \\
        \And
        Stéphane Vialle\\
        Laboratoire Interdisciplinaire des Sciences du Numérique\\ CentraleSupélec, CNRS, Université Paris-Saclay\\
        rue du Belvédère, 91405 Orsay, France\\
        \texttt{stephane.vialle@centralesupelec.fr}
}
\date{\today}

\begin{document}

\renewcommand{\refname}{References}
\renewcommand{\bibname}{References}
\maketitle

\begin{abstract}
In this study, we develop a Diffusion Transformer (referred as to DiT1D) for synthesizing realistic earthquake time histories. The DiT1D generates realistic broadband accelerograms (0-30 Hz resolution), constrained at low frequency by 3-dimensional (3D) elastodynamics numerical simulations, ensuring the fulfillment of the minimum observable physics. The DiT1D architecture, successfully adopted in super-resolution image generation, is trained on recorded single-station 3-components (3C) accelerograms. Thanks to Multi-Head Cross-Attention (MHCA) layers, we guide the DiT1D inference by enforcing the low-frequency part of the accelerogram spectrum into it. The DiT1D learns the low-to-high frequency map from the recorded accelerograms, duly normalized, and successfully transfer it to synthetic time histories. The latter are low-frequency by nature, because of the lack of knowledge on the underground structure of the Earth, demanded to fully calibrate the numerical model. We developed a CNN-LSTM lightweight network in conjunction with the DiT1D, so to predict the peak amplitude of the broadband signal from its low-pass-filtered counterpart, and rescale the normalized accelerograms rendered by the DiT1D. Despite the DiT1D being agnostic to any earthquake event peculiarities (magnitude, site conditions, etc.), it showcases remarkable zero-shot prediction realism when applied to the output of validated earthquake simulations. The generated time histories are viable input accelerograms for earthquake-resistant structural design and the pre-trained DiT1D holds a huge potential to integrate full-scale fault-to-structure digital twins of earthquake-prone regions.
\end{abstract}





\section{Introduction}
\label{sec-Introduction}
Modern high-performance computing tools have been developed in the recent past to numerically solve the elastodynamic problem in complex setups. Such tools are continuously optimized and deployed on different hardware architectures to achieve superior performance on multi-CPU/GPU clusters~\citep{Kirk_Peterson_Stogner_Carey_2006,Kaczmarczyk_et_al_2020,Anderson_et_al_2021,Baratta_et_al_2023}. One of the most challenging  applications of such top-notch software tools (in terms of domain geometry, material properties, source characteristics, etc.) is seismology where, in spite of huge amount of available recordings worldwide, numerical models suffer from poor resolution compared to the intricacy of the highly uncertain Earth's crust structure and seismotectonics configuration. Physics-based approaches to reproduce earthquake scenarios require the simulation of a large multi-scale problem, covering large domain including active faults and spanning across urban regions $\approx$100 km $\times$ 100 km $\times$ 50 km large. Ideally, the minimum resolution of the computational model should be of the order of the meter. State-of-the-art numerical tools are nowadays capable to accurately reproduce the complete earthquake physics. They include complex active fault rupture mechanisms~\citep{Wirp_Gabriel_Ulrich_Lorito_2024}, as well as simulate the wave propagation through intricate geological structures~\citep{Gatti_et_al_2017b,CastroCruz_Gatti_LopezCaballero_2021}. Moreover, such advanced pieces of software simulate the non-linear interaction between the incoming seismic wave field and the impinged shallow soft soil sediments~\citep{DeMartin_Chaljub_Thierry_Sochala_Dupros_Maufroy_Hadri_Benaichouche_Hollender_2021}, as well as the amplifications and surface wave generation induced by underground and surface topography~\citep{Soto_Lopez-Caballero_2024}. Given the size and the details of the models at stake, regional earthquake digital twins are featured by billions of parameters that require careful calibration~\citep{Chaljub_et_al_2010,Maufroy_et_al_2015,Maufroy_et_al_2016,Fu_et_al_2017,Poursartip_et_al_2020,Pitarka_Graves_Irikura_Miyakoshi_Wu_Kawase_Rodgers_McCallen_2021,Touhami_et_al_2022,Smerzini_et_al_2023,Otsuka_Tamari_Fujita_Ichimura_2024,Rojas_et_al_2024}. \\ Nonetheless, end-to-end earthquake scenarios are still hardly achievable. Very rare are so far the examples of complete fault-to-structure analysis, including strong Soil-Structure Interaction (SSI) coupling~\cite[see for instance][]{Hori_Ichimura_2006,Ichimura_et_al_2012,Ichimura_et_al_2017b,Zuchowski_et_al_2018,McCallen_et_al_2020_II}. This is due to the fact that the vibrational fingerprint of critical above-~ground structures and related non-structural system components (such as piping systems and other equipments) is characterized by resonant eigenmodes way above 5 Hz~\citep{Korres_et_al_2023} and up to 40 Hz, whereas physics-based simulations (PBSs) are restricted to a relatively low frequency interval~\citep[reaching up to a maximum frequency between 1 Hz and 10 Hz,][]{Gatti_et_al_2018e}. The reasons behind this unresolved incompatibility between fault-to-site and site-to-structure simulations are two, namely:
\bi
\item[$\bullet$] the increasing required computational demand, implied by the finest spatial discretization that ensures non-aliasing effects;
\item[$\bullet$] the large uncertainty on the underground geological and seismotectonic configuration~\citep[the mechanical properties of huge chunks of Earth's crust, including shallow soil non-linearity,][]{Gatti_et_al_2017b}, the geometry of the geological interfaces between layers~\citep{CastroCruz_et_al_2021a,Lehmann_et_al_2024a} and the geometry of active fault segments, their roughness, their properties, and the tectonic stress acting on them before the earthquake occurs. The main challenge resides in yielding synthetic signals that are both physically consistent and suitable for structural analysis and design in a broadband frequency range~\citep{CastroCruz_Gatti_LopezCaballero_2021,Stupazzini_et_al_2021}.
\ei
In this study, we leverage the power of the Diffusion Transformer (DiT) architecture~\citep{peebles2023scalablediffusionmodelstransformers} to synthesize broadband acceleration time histories based on low-frequency earthquake ground motion numerical simulation. The DiT was conceived as a scalable architecture for image generation, but, hereafter, we propose adapt the DiT architecture to 3C seismic signals. Therefore, since it operates on one-dimensional time histories, we dub the model as DiT1D from now on. The adopted strategy is to learn to generate realistic broadband time histories in a 0-30 Hz frequency band, while preserving the low-frequency part of the spectrum (generated by PBS in blind earthquake forecast), as conditioning. 

\section{Related works}
\label{sec-related_works}
In recent past, deep learning strategies started overcoming the above-~mentioned limitations. Several studies~\citep[see, for instance, ][]{donahue_adversarial_2019,Okazaki_et_al_2019,Okazaki_et_al_2021a,Okazaki_et_al_2021b} showcased the potential of ML in seismic signals generation, with promising perspectives due to both a good quality of the rendered signal and an overall accessible computational costs (considering offline training of the algorithm and online inference cost). Hybrid approaches, mixing numerical simulations and deep learning were concurrently developed, so to better condition the generation with the earthquake physics. In this context, the seminal work by~\citet{Paolucci_et_al_2018} presented a 2-layers feed-forward neural network called ANN2BB, trained to ingest physics-based simulations (PBS) and to predict 10 to 15 pseudo-spectral acceleration values $Sa(f)$ at natural frequencies $f$ below 1.3 Hz. $Sa(f)$ represents the maximum absolute acceleration $Sa$ of a single-degree of freedom oscillator of natural frequency $f$, subjected to the input earthquake. ANN2BB relies on the implicit assumption that the low-to-high frequency mapping could be learned in a supervised way, directly from recorded data, duly low-pass-filtered. Despite the capability of improving ground motion numerical simulations above the original 1.5 Hz of accuracy in terms of amplitude (see, for instance, a soil-structure interaction analysis leveraging ANN2BB, performed by~\citet{Gatti_et_al_2018e}). \citet{Paolucci_et_al_2021} created a validated set of hybrid simulations of past-earthquakes, called BB-SPEEDset, adopting ANN2BB. However, ANN2BB struggles in reproducing the low-to-high frequency mapping in the phase space (random phases are added at high-frequency) because of the intrinsic definition of the pseudo-spectral acceleration $Sa$. \\

More recently, several approaches sprouted from the seminal idea informing ANN2BB, most of them adopting variations of Generative Adversarial Neural Networks~\citep[GANs,][]{goodfellow_generative_2014}. The first attempt was done by~\citet{Gatti_Clouteau_2020}, who constructed two adversarial auto-encoders, one for low-frequency and one for broadband signals encoding, with a third adversarial auto-encoder mapping the corresponding low-frequency latent code into the broadband one. The authors performed signal-to-signal translation in the latent space and the broadband decoder used to decode broadband time histories, conditioning the low-frequency part by a numerical simulation. Alternative variations to the original formulation can be found in~\citet{PhD_Jacquet_2024}. Customized conditional GANs were conceived by \citet{Florez_et_al_2022} and by~\citet{Esfahani_et_al_2023} to simulate non-stationary ground-motion recordings conditioned by metadata such as the earthquake moment magnitude, source-to-site distance, harmonic average of shear-wave velocity of the first 30 m down deep (the so called $V_{S,30}$, widely adopted in seismology as a proxy for the site conditions). While \citet{Florez_et_al_2022} predicted the whole 3-component (3C -- east-west, north-south, up-down) normalized time series and the peak ground amplitude, ~\citet{Esfahani_et_al_2023} adopted a computer-vision framework inspired by~\cite{Ho_et_al_2020,saharia_image_2021,copet_simple_2023}. Other noticeable studies involving combinations of conditional GANs have been realized by~\citet{pascual2017gan,Yang_Wang_Chi_Feng_2022, cao2022gan,Li_Yoon_Ku_Ko_2024,Chen_Pan_Zhang_Li_2024,Neelamraju_Basu_Raghukanth_2024,Shen_Ni_Ding_Xiong_Zhong_Chen_2024, Masoudifar_Mahsuli_Taciroglu_2025} among others. \citet{Aquib_Mai_2024} used GAN in conjunction with two Fourier Neural Operators (FNO) to generate broadband ground motion accelerograms and establishing a relationship between normalized low‐pass filtered and broadband waveforms in time and frequency.\\

However, GANs have shown major limitations in practical uses, because of the slow training convergence~\citep{barnett_convergence_2018} and \textit{mode collapse}~\citep{thanh-tung_catastrophic_2020}. As an alternative to GANs, Variational Autoencoders (VAEs) have been proven successful to generate realistic ground motion synthetics across large geographic regions~\citep{Ren_et_al_2024}. More recently, denoising diffusion probabilistic models~\citep[DDPM,][]{Ho_et_al_2020} have emerged as superior alternatives to GANs and VAEs~\citep{Vivekananthan_2024}. Diffusion models, particularly those employing Gaussian or cold diffusion~\citep[which replaces Gaussian noise models with alternative degradation processes,][]{bansal2022cold}, offer advantages such as flexibility, reconstruction quality, and robustness against complex noise. The use of diffusion models has demonstrated significant success in generating realistic time series data as well~\citep{lee_nu-wave_2021,han_nu-wave_2022}. Cold Diffusion effectively restores signals impaired by diverse degradation, demonstrating its potential in speech source separation under non-Gaussian noise. Moreover, the pioneering work of~\citet{durall2023seismic,Trappolini_et_al_2024} applied Gaussian-based and cold diffusion models respectively, to denoise either seismic images or strong motion time histories. Diffusion models are being progressively used to generate realistic ground motion scenarios, conditioned by metadata related to the site and to the event characteristics~\citep[magnitude, source-to-site distance, source and site coordinates, ]{Bergmeister_Palgunadi_Bosisio_Ermert_Koroni_Perraudin_Dirmeier_Meier_2024,Huang_Sun_You_Yang_Lu_Zhang_2025}. For instance, \citet{Bi_Nakata_Nakata_Ren_Wu_Mahoney_2025} developed a multi-conditional diffusion U-shaped transformer for the prediction of earthquake ground motion spectral amplitudes up to 15 Hz, for arbitrary source-station configurations. The authors conditioned the transformer learning on stations and source's coordinates, other than the traditional variables such as earthquake magnitude. An adaptive Phase Retrieval Method (PRM) is designed to reconstruct the phase spectrum and produce the final waveform.

\subsection*{Our contribution}
\label{subsec-Our_contribution}
The innovative idea of the DiT1D is to replace the standard backbone architecture adopted in denoising diffusion models, \textit{i.e.,} the U-Net~\citep{Ronneberger_Fischer_Brox_2015,Ho_et_al_2020}. Adopting the guidance strategy proposed in the original DiT paper~\citep{peebles2023scalablediffusionmodelstransformers}, consisting in the use of multi-head cross attention layers, the proposed DiT1D enhances low-frequency simulations and generate stochastic broadband earthquake accelerograms. The DiT1D learns the low-to-high frequency mapping, guided by the low-pass filtered input recordings. Contrary to similar studies involving diffusion models conditioned by event and site characteristics~\citep{Bi_Nakata_Nakata_Ren_Wu_Mahoney_2025}, our DiT1D renders single-stations broadband accelerograms, increasing the realism of numerical simulations up to 30 Hz. Moreover, DiT1D does not need to learn amplitude and phase separately~\citep[such as done by ][]{Esfahani_et_al_2023,Bi_Nakata_Nakata_Ren_Wu_Mahoney_2025} and its training procedure makes it agnostic to site and scenario specific conditions. Thus, the DiT1D generalizes better and can be applied zero-shot to enhance the realism of the conditioning time histories, generated by high-performance numerical solvers. \\ The major contributions of this study are the following:
\bi
    \item[$\bullet$] We propose a scalable diffusion transformer inspired by~\citet{peebles2023scalablediffusionmodelstransformers} and adapted to earthquake seismograms, for physically-guided generation of broadband (0-30 Hz) seismic signals.

    \item[$\bullet$] At training time, the low-frequency portion of the input signal is used as guidance in for the DiT1D inference. Multi-Head Cross-Attention (MHCA) layers are used to guide the DiT1D inference. 

    \item[$\bullet$] The trained DiT1D reaches state-of-art performance as stochastic earthquake generator, conditioned by cheap yet realistic numerical simulations, to preserve and to incorporate the minimum knowledge on the earthquake physics.

    \item[$\bullet$] We trained a lightweight network alongside the DiT1D and on the same database, in a supervised way. Its goal is to infer the peak value of the broadband signals, based on the low-frequency part, and then use it to scale the normalized time history yielded by the DiT1D, according to the physics of the earthquake.

    \item[$\bullet$] We perform a rigorous study on the DiT1D performance with the respect to its scalability on multiple GPUs (Graphics Processing Units). The analysis focuses on training and inference steps, duly adjusted to reduce the overall computation burden and provide a portable AI surrogate model for earthquake ground motion prediction.

    \item[$\bullet$] We tested the DiT1D on a database of earthquake numerical simulations~\citep[namely, the BB-SPEEDset created by][]{Paolucci_et_al_2021}, considering historical seismic events employed to validate the numerical model. The results prove that the DiT1D can generate realistic broadband time histories, preserving the low-frequency/physics-based part of the spectrum, while accurately reproducing the unseen high-frequency components. 
\ei

\Cref{sec-Method} introduces the diffusion process and the DiT1D architecture adopted in the study. In \Cref{sec-Data}, we describe the preprocessing pipeline applied to the training dataset and the recorded and synthetic databases employed to test the DiT1D's zero-shot performance. \Cref{sec-Results} discusses the inference performance of the DiT1D, as well as a fair comparison with state-of-the-art neural architectures for time series generation. Finally, \Cref{sec-Conclusions} concludes the study and discusses future directions.

\section{Method}
\label{sec-Method}
\subsection{Denoising Diffusion Process}
\label{sec-diffusionprocess}
\begin{figure}[ht!]
    \centering
    \includegraphics[width=\textwidth]{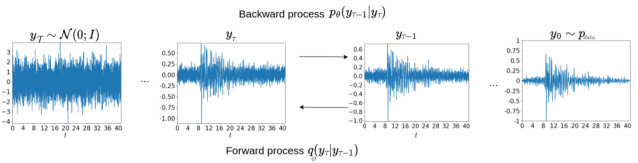}
    \caption{Scheme of the diffusion processes for a seismic signal $\yv[0]\ptp\sim p_{\text{data}}$. The \emph{forward} Markov process blurs $\yv[0]$ towards white noise sample $\yv[\mathcal{T}]\sim \mathcal{N}\pdp{\zerov,\ID}$ for large $\mathcal{T}$. The kernel proxy associated to the \emph{forward} Markov process reads $q_{\phiv}(\yv[\tau] | \yv[\tau-1])$.}
    \label{fig-diffusionprocess}
\end{figure}

Diffusion models generate samples with a slowly decaying energy spectrum, efficiently taking advantage of priors~\citep{Oommen_Bora_Zhang_Karniadakis_2024}. In this work, we trained a diffusion model -- conditioned on the low-frequency part of the spectrum of an earthquake signal -- to reconstruct broadband realistic accelerograms.\\ In particular, we adopted state-of-the-art denoising diffusion model~\citet{Ho_et_al_2020,song_denoising_2022}. The latter adopts a discrete Monte Carlo Markov chain sampling to synthesize new instances according to the unknown data distribution $p_0$, starting from a reference one $p_{\text{ref}}$~\citep{Ho_et_al_2020}. The generation consists into two discrete diffusion processes, corresponding to two Markov chains in a discrete setup. We refer to real data as the 3C (EW: East-West, NS: North-South, US: Up-Down) broadband seismograms $\yv[0]:\R_{\ge 0}\ni t\mapsto \psp{y^{EW}\ptp,y^{NS}\ptp,y^{UD}\ptp}^T\in\R^3$. Omitting the dependence on time $t$ for the sake of conciseness, the \emph{forward} chain (or \emph{forward} process) consists into an iterative generation of progressively blurred 3C samples $\pbp{\yv[\tau]}_{\tau = 1}^{\mathcal{T}}$. The latter sequence is a Markov chain obtained by adding white noise to the original data $\yv[0]$ (see \Cref{fig-diffusionprocess}). $\mathcal{T}\in \N$ represents the duration of the \emph{forward} diffusion process (the total number of blurring steps).
To blur original data, the forward process has a scheduled upstream noise variance $(\beta_{\tau})_{\tau=1}^{\mathcal{T}}$. Following~\citet{Ho_et_al_2020}, $\beta_{\tau}$ is incremented linearly from 0.0001 to 0.02, and the Gaussian transition kernel iteratively generates new instances $\pbp{\yv[\tau]}_{\tau = 1}^{\mathcal{T}}$, according to the following expression:
\be
q(\yv[\tau] | \yv[\tau-1]) := \mathcal{N}(\yv[\tau];\yv[\tau-1]\sqrt{\alpha_{\tau}},\pdp{1-\alpha_{\tau}}\ID)
\label{eq-qtt1}
\ee
Thanks to the Markov property, $\yv[\mathcal{T}]$ is sampled according to the following conditional distribution:
\be
q(\yv[\mathcal{T}]| \yv[0]) := \mathcal{N}(\yv[\mathcal{T}];\sqrt{\bar{\alpha_{\mathcal{T}}}}\yv[0], (1-\bar{\alpha_{\mathcal{T}}})\ID)
\label{eq-qT0}
\ee
with $\bar\alpha_{\tau} =\prod_{s=1}^{\tau}\alpha_s$. $\bar{\alpha}_{\mathcal{T}}\to 0$ if $\mathcal{T}\to +\infty$, where for $\mathcal{T}$ sufficiently large~\citep[it is usually set to 1000, according to][]{Ho_et_al_2020}, $q(\yv[\mathcal{T}]| \yv[0])$ exponentially converges to the \emph{reference} probability density $p_{\text{ref}} = \mathcal{N}(\zerov,\ID)$. One notices that the conditional distribution can be conditioned on $\yv[0]$ according to the Bayes theorem~\citep{Chan_2025}:
\be
q\pdp{\yv[\tau-1]\vert\yv[\tau],\yv[0]}=\frac{ q\pdp{\yv[\tau]\vert\yv[\tau-1],\yv[0]} q\pdp{\yv[\tau-1]\vert\yv[0]} }{q\pdp{\yv[\tau]\vert\yv[0]}}=\mathcal{N}\pdp{\yv[\tau-1]; f_{\tau}\yv[\tau]+g_{\tau}\yv[0],\sigma_{\tau}^2\ID }
\label{eq-bayes-y0}
\ee
with $f_{\tau}=\frac{\pdp{1-\bar{\alpha}_{\tau-1}}\sqrt{\alpha_{\tau}}}{1-\bar{\alpha}_{\tau}}$, $g_{\tau}=\frac{\pdp{1-\alpha_{\tau}}\sqrt{\bar{\alpha}_{\tau-1}}}{1-\bar{\alpha}_{\tau}}$ and $\sigma_{\tau}^2=\frac{\pdp{1-\alpha_{\tau}}\pdp{1-\sqrt{\bar{\alpha}_{\tau-1}}}}{1-\bar{\alpha}_{\tau}}$. To generate realistic signals $\yhv\sim p_{\text{data}}$, one must \emph{reverse} the above-mentioned Markov chain, \emph{i.e.}, replacing $\tau$ with $\mathcal{T}-\tau$ and denoising the sample $\yv[\mathcal{T}]\sim p_{\text{ref}}$. In other words, from a white noise sample we yield a coherent seismic signal $\yhv \sim p_{\text{data}}$. Along the discrete reverse process, the diffusion model estimates the transition probability $p(\yv[\tau-1] | \yv[\tau])$. ~\citet{Ho_et_al_2020} proposed to parametrize the \emph{backward} transition kernel according to the following expression:
\be
p_{\thetav}(\yv[\tau-1] | \yv[\tau]) := \mathcal{N}(\yv[\tau-1];\muv[\thetav](\yv[\tau];\tau),\sigma_{\tau}^{2}\ID)
\label{eq-eq3}
\ee
with $p_{\thetav}(\yv[0]|\yv[\mathcal{T}]) := p(\yv[\mathcal{T}]) \prod_{t=1}^{T-1} p_{\thetav}(\yv[\tau-1]|\yv[\tau])$. $\muv[\thetav](\yv[\tau];\tau)$ acts as a \emph{proxy} of the mean of the transition kernel $q(\yv[\tau-1] | \yv[\tau],\yv[0])$ expressed in \Cref{eq-bayes-y0}. In other words, $\muv[\thetav]$ is designed to match the mean of the forward process $f_{\tau}\yv[\tau]+g_{\tau}\yv[0]$. By replacing $\yv[0]$ with $\frac{\yv[\tau]-\sqrt{1-\bar{\alpha}_{\tau}}\boldsymbol{\epsilon}_0}{\sqrt{\bar{\alpha}_{\tau}}}$ as per \Cref{eq-qT0} (with $\boldsymbol{\epsilon}_0\sim\mathcal{N}\pdp{\zerov,\ID}$) the proxy mean can be rewritten as:
\be
\muv[\thetav](\yv[\tau];\tau)=\frac{1}{\sqrt{\alpha_{\tau}}}\cdot\left( \yv[\tau] - \frac{1 - \alpha_{\tau}}{\sqrt{1 - \bar{\alpha_{\tau}}}}\NN[\thetav]\pdp{\yv[\tau];\tau}\right)
\label{eq-NNtheta}
\ee
where $\NN[\thetav](\yv[\tau];\tau)$ represents a neural network (the DiT1D) progressively denoising $\yv[\tau]$ by a factor $\frac{1 - \alpha_{\tau}}{\sqrt{1 - \bar{\alpha_{\tau}}}}\NN[\thetav]\pdp{\yv[\tau];\tau}$ and sample $\yv[\tau-1]$, according to the following expression:
\be
\yv[\tau-1] = \frac{1}{\sqrt{\alpha_{\tau}}}\cdot\left( \yv[\tau] - \frac{1 - \alpha_{\tau}}{\sqrt{1 - \bar{\alpha_{\tau}}}}\NN[\thetav]\pdp{\yv[\tau];\tau}\right)+\sigma_t\boldsymbol{\epsilon},\quad \boldsymbol{\epsilon} \sim \mathcal{N}\pdp{\zerov,\ID}
\label{eq-eq4}
\ee
The iterative process in \Cref{eq-eq4} converges towards a mean estimate of the uncorrupted initial data $\yv[0]$, from now on referred as to $\yhv$. At inference time, the backward process is employed to sample new data $\yhv$, starting from $\yv[\mathcal{T}]\sim\mathcal{N}\pdp{\zerov,\ID}$. In this study, following~\citet{Rombach_Blattmann_Lorenz_Esser_Ommer_2022}, the low-frequency signal  obtained from physic-based simulation (pbs) $\xv\sim p_{\text{pbs}}$ guides the generation. The adopted conditioning approach corresponds to the \emph{classifier-free} guidance proposed by \citet{Ho_Salimans_2022}, with guidance scale of 1. In practice, $\NN[\thetav]$ is equipped with MHCA layers~\citep{Vaswani_Shazeer_Parmar_Uszkoreit_Jones_Gomez_Kaiser_Polosukhin_2023} that statistically align the denoised sample $\yv[\tau-1]$ with $\xv$, guiding the denoiser $\NN[\thetav]\pdp{\yv[\tau],\xv;\tau}$ at low frequency (see \Cref{subsec-Model_Architecture}). This approach is similar to auto-regression in frequency space~\citep{Oommen_Bora_Zhang_Karniadakis_2024}. With conditioning, the parametric Markov transition probability corresponds to $ p_{\thetav}(\yv[\tau-1] | \yv[\tau],\xv)$.

\noindent
The training process aims at maximizing the \emph{Evidence} \emph{Lower} \emph{Bound} \citep[ELBO,][]{Kingma_Welling_2014}. Since we schedule the variance $\sigma^{2}_{\tau}$ of the Gaussian transition kernel in \Cref{eq-eq3}, the ELBO loss simplifies to the following expression~\citep{Ho_et_al_2020,Rombach_Blattmann_Lorenz_Esser_Ommer_2022,Chan_2025}:
{\scriptsize
\be
\Loss_{\text{ELBO}}\pdp{\thetav}=\emean[\tau\sim \mathcal{U}\pdp{\llbracket 1,\mathcal{T}\rrbracket},\xv\sim p_{\text{pbs}},\yv[0]\sim p_{\text{data}},\vct{\epsilon}{}{}\sim\mathcal{N}\pdp{\zerov,\ID}]\left[\Big\Vert\vct{\epsilon}{\tau}{} - \NN[\thetav]\left(\sqrt{\bar{\alpha_{\tau}}}\yv[0] + \sqrt{1- \bar{\alpha_{\tau}}}\vct{\epsilon}{}{},\xv;\tau\right) \Big\Vert^{2}\right]
\label{eq-eq5}
\ee}
Computing the mean in \Cref{eq-eq5} over $\xv\sim p_{\text{pbs}}$ and $\yv[0]\sim p_{\text{data}}$ necessitates a high-fidelity pbs for each earthquake time history. To bypass the inherent prohibitive time and computational costs, we sample the tuple $\pdp{\xv[0],\yv[0]}$, with $\xv[0]$ representing the low-pass filtered version of $\yv[0]$. This strong assumption resembles to a data masking~\citep{He_Chen_Xie_Li_Dollar_Girshick_2022}, but in the frequency domain. Therefore, \Cref{eq-eq5} simplifies to:
{\scriptsize
\be
\Loss_{\text{ELBO}}\pdp{\thetav}=\emean[\tau\sim \mathcal{U}\pdp{\psp{1,\mathcal{T}}},\pdp{\xv[0],\yv[0]}\sim p_{\text{data}},\vct{\epsilon}{}{}\sim\mathcal{N}\pdp{\zerov,\ID}]\left[\Big\Vert\vct{\epsilon}{\tau}{} - \NN[\thetav]\left(\sqrt{\bar{\alpha_{\tau}}}\yv[0] + \sqrt{1- \bar{\alpha_{\tau}}}\vct{\epsilon}{}{},\xv[0];\tau\right) \Big\Vert^{2}\right]
\label{eq-eq5bis}
\ee
}
To further guide the conditional denoiser, the ELBO loss in \Cref{eq-eq5bis} is penalized by the L2 norm of the high-frequency residual $\rv\psp{\yv[\tau],\yv[0]}\ptp = \mathcal{F}_{f\ge 10 Hz}\pdp{\yv[\tau]\ptp-\yv[0]\ptp}$, where $\mathcal{F}_{f\ge 10 Hz}$ is the high-pass Butterworth's filter for frequencies above 10 Hz. The latter frequency value has been arbitrary chosen after some trial and error procedure. Therefore, the final loss function reads:
\be
\Loss\pdp{\thetav}=\Loss_{\text{ELBO}}\pdp{\thetav}+\lambda\mathcal{R}\pdp{\thetav}
\ee
with the residual penalty:
\be
\mathcal{R}\pdp{\thetav}=\emean[\tau\sim \mathcal{U}\pdp{\psp{1,\mathcal{T}}},\pdp{\xv[0],\yv[0]}\sim p_{\text{data}},\vct{\epsilon}{\tau}{}\sim\mathcal{N}\pdp{\zerov,\ID}]\psp{\norm{\rv\pdp{\yv[\tau],\xv[0]}}_{L^2}^2}
\ee
where the L2 norm is computed in the Fourier's space, by exploiting the Parseval's identity. After some tuning, we chose a $\lambda$ penalty coefficient of 0.01.

\noindent
The backward Markov chain expressed in \Cref{eq-eq4} requires as many steps as the forward one (1000 in this case) to infer new samples $\yhv$. Therefore, to reduce inference time, we adopted a variant of diffusion model called Denoising Diffusion Implicit Models~\citep[DDIM,][]{song_denoising_2022}. DDIM is based on a non-Markovian backward process, expressed in \Cref{eq-ddim}, that renders new samples in far fewer steps ($\approx$100 steps in this work):
\be
\begin{split}
    \yv[\tau-1] & = \sqrt{\frac{\alpha_{\tau-1}}{\alpha_{\tau}}}\pdp{\yv[\tau]-\sqrt{1-\alpha_{\tau}}\NN[\thetav](\yv[\tau],\xv;\tau)}+ \\
                & +
    \sqrt{1-\alpha_{t-1} - \beta_{\tau}}\NN[\thetav](\yv[\tau],\xv)
    +
    \sqrt{\beta_{\tau}}\vct{\epsilon}{}{}, \\ &\vct{\epsilon}{}{} \sim \mathcal{N}(\zerov,\ID)
\end{split}
\label{eq-ddim}
\ee
with $\beta_{\tau}=1-\alpha_{\tau}$. In \Cref{eq-ddim}, the term $\sqrt{\frac{\alpha_{\tau-1}}{\alpha_{\tau}}}\pdp{\yv[\tau]-\sqrt{1-\alpha_{\tau}}\NN[\thetav](\yv[\tau],\xv;\tau)}$ represents the estimate of the original data $\yv[0]$ (conditioned by physics-based signal $\xv$). The term $\sqrt{1-\alpha_{t-1} - \beta_{\tau}}\NN[\thetav](\yv[\tau],\xv;\tau)$ acts as a weighting coefficient that scales the predicted noise direction. The DDIM is carefully designed to maintain the same marginal noise distributions of the original Denoising Diffusion Probabilistic Model~\citep[DDPM,][]{Ho_et_al_2020} in \Cref{eq-eq4} at each time step, but the backward process converges faster than the original DDPM. Moreover, DDIM holds another practical advantage: a model trained using the original DDPM loss function in \Cref{eq-eq5bis} can be used to sample in a DDIM fashion (see \Cref{eq-ddim}) at inference time, with no need for retraining.
\subsection{Model Architectures}
\label{subsec-Model_Architecture}
In this study, the $\NN[\thetav]$ neural network in transformer tailored for time histories. The DiT1D takes as input the 3C earthquake accelerograms $\yv[0]\ptp$ and the condition $\xv[0]\ptp$ at training time, and $\xv\ptp$ at inference time. Both $\yv[0]\ptp$ and $\xv[0]\ptp$ are normalized by their respective maximum absolute value (corresponding to the so called Peak Ground Amplitude, PGA). To preserve the accelerogram amplitude, a lightweight architecture based on Convolutional Neural Networks~\citep{Ajit_et_al_2020} and Long Short-Time Memory blocks~\citep{Hochreiter_Schmidhuber_1997} is trained to predict the PGA of the broadband signal $\yv[0]$, based on unnormalized conditioning signal $\xv[0]$ and normalized denoised signal $\yhv$. We trained such a CNN-LSTM (depicted in \Cref{subsec-Amplitude_predictor_model}), in a supervised way, so to learn the mapping $\pdp{\xv[0],\yhv}\mapsto \underset{t}{\max}\vert \yv[0]\ptp\vert$. In such a way, we design the CNN-LSTM network to infer the relative amplitude scaling between low-frequency signal $\xv[0]\ptp$ and broadband one $\yv[0]\ptp$, based on the normalized prediction of the DiT1D $\yhv\ptp$.

\subsubsection{Transformer architecture}
\label{subsec-Diffusion_Transformer_Architecture}
The DiT1D is an adaptation to 3C time histories of the original Diffusion Transformer (DiT) proposed by~\citet{peebles2023scalablediffusionmodelstransformers} for image generation. The DiT1D architecture builds upon the classical transformer~\citep{Vaswani_Shazeer_Parmar_Uszkoreit_Jones_Gomez_Kaiser_Polosukhin_2023}. Traditionally, diffusion models are based on a backbone UNet architecture~\citep{Ronneberger_Fischer_Brox_2015}, but the DiT architecture successfully showcased state-of-the-art performance in image generation~\citep{Hatamizadeh_et_al_2024}. To  the best of our knowledge, the DiT has not yet been tested in audio super-resolution problems with conditioning.

\begin{figure}[!ht]
     \centerline{\includegraphics[width=0.9\textwidth]{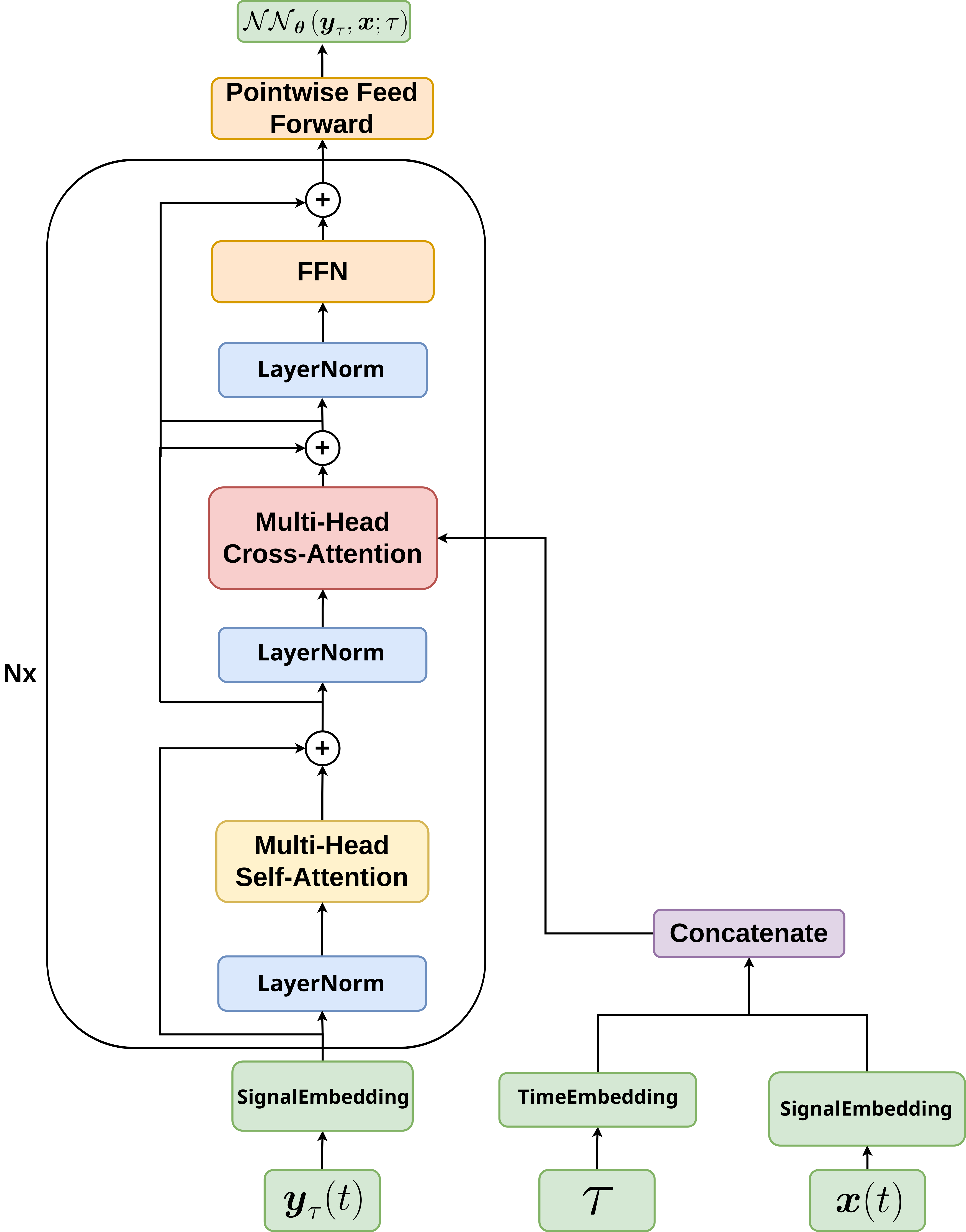}}
     \caption{DiT1D architecture, based on the original architecture proposed by~\citet{peebles2023scalablediffusionmodelstransformers}.}
     \label{fig-diffusion transformer}
\end{figure}

Our implementation introduces conditioning through Multi-Head Cross-Attention (MHCA) blocks, as illustrated in \Cref{fig-diffusion transformer}.
After hyper-parameter optimization, the best model configuration consists into a $N_x=12$ layers architecture, with hidden embedding of dimension 1024, a patch size of 8, and 16 attention heads each. The transformer's head is Point-Wise Feed-Forward Network (FFN), consisting into two fully-connected Multi-Layer Perceptrons (MLPs). We employed two distinct architectures to encode the condition $\xv[0](t)$ (or, at inference time, $\xv(t)$) and the diffusion step $\tau$, respectively. The \texttt{SignalEmbedding} contains four residual blocks, each containing 1D convolutional layers followed by Layer Normalization~\citep[\texttt{LayerNorm} in \Cref{fig-diffusion transformer}, see][]{Ba_Kiros_Hinton_2016}. 
The \texttt{TimeEmbedding} block uses sine/cosine temporal encoding, to ensure a frequency-based representation, followed by a two-layer fully-connected MLP. 

\subsubsection{Amplitude predictor model}
\label{subsec-Amplitude_predictor_model}
\begin{figure}[ht!]
     \centerline{\includegraphics[width=0.4\textwidth]{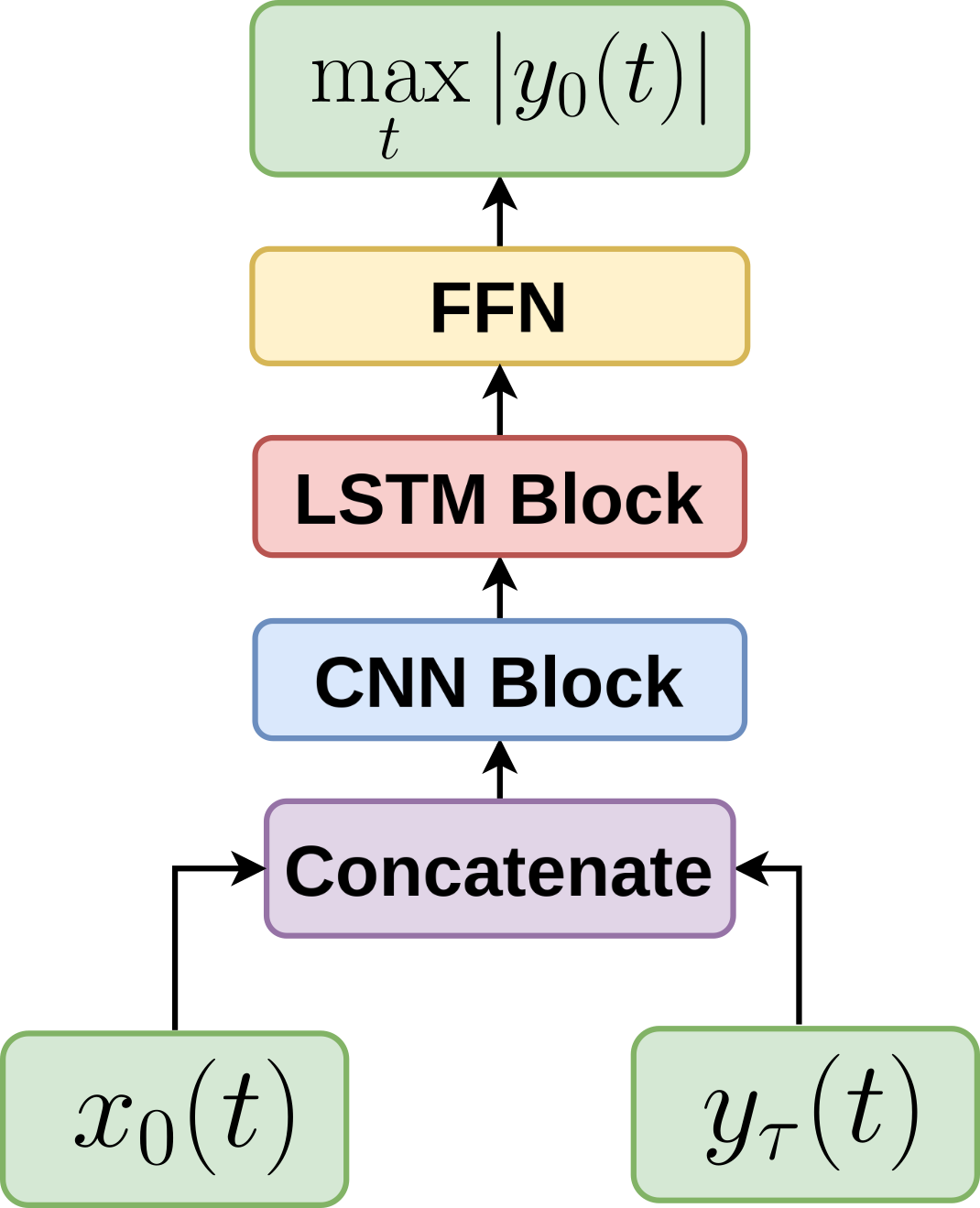}}
     \caption{CNN-LSTM architecture for predicting $\yv\ptp$ maximum absolute value, from }
     \label{fig-CNNLSTM}
\end{figure}

To generate realistic seismic signals, we also implemented a CNN-LSTM capable of inferring broadband signal's maximum absolute amplitude (also called, in earthquake engineering, Peak Ground Acceleration, or PGA) from $\xv[0]$ (or $\xv\ptp$ at inference time). At training time, the input to the lightweight CNN-LSTM (depicted in \Cref{fig-CNNLSTM}) is the couple $\pdp{\xv[0],\yhv}$ and predicts $\yv[0]$ PGA.
The CNN block extracts equivariant temporal features from the 3C time histories. CNN blocks have an increasing filter depth (32$\to$64$\to$128$\to$256) and kernel size of 3, with ReLu activation and batch normalization, efficiently extract hierarchical features from the concatenated low-frequency signal and diffusion-generated output. The LSTM block, instead, captures temporal dependencies in the seismic data and are insensitive to unnormalized inputs, because of the gated mechanism consisting of hyperbolic tangent and sigmoid activation functions~\citep{Mousavi_Zhu_Ellsworth_Beroza_2019}. We introduce two bidirectional LSTM layers (512$\to$256 hidden units) to capture temporal dependencies in both directions, allowing for modeling the relationship between waveform characteristics and PGA outputted by the final FFN layer.
\subsection{Training details and Evaluation Process}
\label{subsec-Training_details_and_Evaluation_Process}
\Cref{alg-training} and \Cref{alg-prediction} summarize training and inference processes respectively.
\begin{figure}[ht!]
        \begin{algorithm}[H]
            \caption{DiT1D training algorithm by stochastic gradient on $N_b$ mini-batches of size $n$ and over $n_e$ epochs.}\label{alg-training}
            \begin{algorithmic}[1]
                \State $i=0$, $\thetav[][(i)]:=\thetav[][(0)]$ \For{$i<n_e$}

                \For{$b\sim \mathcal{U}(\left\lbrace 1,...,N_b\right\rbrace)$}
                \State $\ell=0$
                \For{$j\sim \mathcal{U}(\left\lbrace 1,...,n\right\rbrace)$}


                \State $\xv[0,j]=\mathcal{F}_{f\le 1 Hz}\pdp{\yv[0,j]}$

                \For{$\tau \sim \mathcal{U}(\left\lbrace 1,...,\mathcal{T}\right\rbrace)$}

                \State $\vct{\epsilon}{}{} \sim \mathcal{N}(\zerov,\ID)$

                \State $\ell += \Big\Vert\vct{\epsilon}{}{} - \NN[\thetav]\left(\sqrt{\bar{\alpha_{\tau}}}\yv[0,j] + \sqrt{1- \bar{\alpha_{\tau}}}\vct{\epsilon}{}{},\xv[0,j];\tau\right) \Big\Vert^{2}$

                \EndFor

                \EndFor
                \State $\Loss\left(\thetav[][(i)]\right):+=\frac{\ell}{N_b}$

                \State
                $\thetav[][(i+1)]:=\textit{AdamW}\pdp{\thetav[][(i)],\eta^{(i)}}$
                \EndFor
                \EndFor
            \end{algorithmic}
        \end{algorithm}
        \begin{algorithm}[H]
            \caption{DiT1D sampling algorithm (inference)}\label{alg-prediction}
            \begin{algorithmic}[1]
                \State $\yv[\mathcal{T}] \sim \mathcal{N}(\zerov,\ID)$
                \For{$\tau = \mathcal{T},...,1$}
                \State $\vct{\epsilon}{\tau}{} \sim \mathcal{N}(\zerov,\ID)$ if $\tau > 1$ else $\vct{\epsilon}{\tau}{}=\zerov$
                \State $\yv[\tau-1] = \frac{\yv[\tau]}{\sqrt{\alpha_{\tau}}}+\frac{\sqrt{1-\alpha_{\tau}}}{\sqrt{1-\bar{\alpha_{\tau}}}}\left(  \sqrt{\bar{\alpha}_{\tau}}\boldsymbol{\epsilon}_{\tau} -\sqrt{1-\alpha_{\tau}}\NN[\thetav]\pdp{\yv[\tau],\xv;\tau} \right)$
                \EndFor
            \end{algorithmic}
        \end{algorithm}
\end{figure}
We adopted AdamW optimizer~\citep{loshchilov_decoupled_2019} to train the DiT1D, which has been proven to be
presents better generalization capabilities than Adam~\citep{kingma_adam_2017} with a learning rate of 0.0001 and a weight decay of 0.1. We trained the model for $n_e=$200 epochs, with a batch size of 64
per GPU and distributed data parallel training on 4 GPU Nvidia A100 SXM4 80 Go (available on the Jean Zay supercomputer at IDRIS-GENCI\footnote{\url{http://www.idris.fr/jean-zay}}, France), for a total 20 hours wall-time. When evaluating our model, we first used the DDPM sampling algorithm described in \Cref{sec-diffusionprocess} using 1000 time steps. We adopted the DDIM algorithm~\citep{song_denoising_2022} to reduce the inference time to 100 steps. We applied Exponential Moving Average (EMA) with a forge factor of 0.999, to mitigate stochastic error and denoising instability~\citep{Lu_Zhou_Bao_Chen_Li_Zhu_2022}.

\section{Data}
\label{sec-Data}
\subsection{Training dataset}
\label{subsec-training dataset}
\begin{figure}[!ht]
     \centerline{\includegraphics[width=0.8\textwidth]{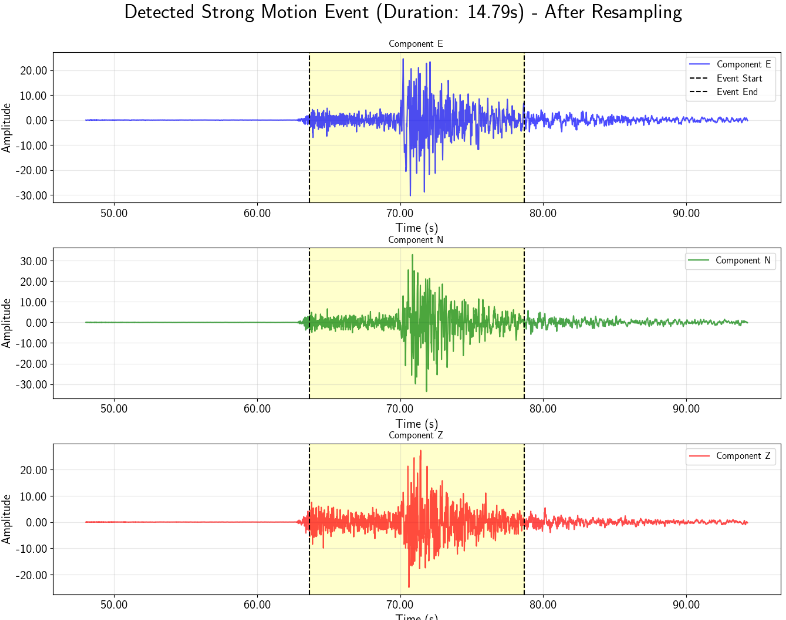}}
     \caption{Example of windowing to create training data, based on the STA/LTA method.}
     \label{fig-detection method}
\end{figure}

The DiT1D must learn the low-to-high frequency mapping on single-station 3C accelerograms. In order to train the DiT1D in a supervised way, we employed 14000 3C seismic signals randomly selected from the Engineering Strong Motion Database (ESM), version 2.0~\citep{Luzi2020ESM}, corresponding to events with magnitude $M_W\ge$4 and maximum epicentral distance of 200 km. To create the paired measurements ($\xv[0]\ptp$,$\yv[0]\ptp$) in our dataset, we performed several processing steps. Firstly, inspired by~\citep{Mousavi_Sheng_Zhu_Beroza_2019}, a 60 s wide time windowing was applied to $\yv[0]\ptp$, adopting either the recursive STA/LTA, either the Z-detector~\citep{Withers_Aster_Young_Beiriger_Harris_Moore_Trujillo_1998} algorithms to determine the window start and end time. This choice implies that each training instance has 6000 time steps, leading to a reasonable computational cost for each training epoch, with an optimum mini-batch size of 32. We used the minimum of all start times and the maximum of all end times calculated by these methods to define the temporal bounds of each seismic event (see \Cref{fig-detection method} for an example).
As shown in \Cref{fig-detection method}, this combination successfully identifies the correct event boundaries. We excluded events whose duration is longer than 60 s. For shorter events, we extracted the signal using the new start and end times and added random noise padding. Each signal $\yv[0]\ptp$ was band-passed filtered between 0.1 and 30 Hz and corrected according to the procedure proposed by~\citet{paoli11}. For training purpose, we low-pass filtered $\yv[0]\ptp$ at 1 Hz, with a Butterworth filter, so to obtain $\xv[0]\ptp=\mathcal{F}_{f\le 1 Hz}\pdp{\yv[0]\ptp}$. The choice of the cutoff frequency aligns with the widely adopted seminal work of~\citet{Graves_Pitarka_2010}, which blends 0-1 Hz numerically simulated accelerograms with stochastic strong motion time history at high-frequency. The choice of 1 Hz cutoff frequency has historical background, since in the early era of numerical simulations, the limited availability of high-performance computing resources were capping the resolution of 3D regional earthquake computer simulations to low frequency range. A cutoff frequency of 1 Hz  is also compatible with the available source models obtained from traditional seismic source inversion~\citep{Zhu2022}.

While this sample size is relatively modest compared to similar studies in the field, it provided an ideal test case to evaluate DiT1D's generative capabilities on a low-to-medium-sized dataset.
Therefore, we performed data augmentation to enlarge the database, to prevent overfitting, and to force the DiT1D to learn complex patterns and invariance features from the input signals themselves~\citep{ZHU2020151}. Following~\citet{Masoudifar_Mahsuli_Taciroglu_2025}, we leveraged full random time-shift, in order for the model to not learn any bias on the time of the P-wave arrival. We also augment the dataset by random rotation of horizontal components. However, we did not perform any noise augmentation, as it interfered with the DiT1D training.

\subsection{Datasets to test zero-shot inference}
\label{subsec-datasets_to_test_zero-shot_inference}
To evaluate the DiT1D generalization capability, in \Cref{subsec-Zero_shot_capability} we present the results of the tests conducted on its zero-shot inference performance on two different datasets. First, we analyze the Goodness-of-Fit (GOF) generated from 500 seismic signals derived from the STanford EArthquake Dataset (STEAD) dataset~\citep{Mousavi_Sheng_Zhu_Beroza_2019}. STEAD is a curated set of earthquake strong motion accelerograms recorded worldwide, and it is employed to test DiT1D zero-shot capability on recorded data $\yv[0]$, conditioned by low-pass-filtered counterpart $\xv[0]$. Second, we conduct a single-station study, comparing the super-resolved signals generated using the simulated low-frequency time histories $\xv$, obtained from the BB-SPEEDset dataset~\citep{Paolucci_et_al_2021}. BB-SPEEDset is a broadband (0-30 Hz) earthquake ground motions dataset, containing synthetic accelerograms obtained from the physics-based numerical simulation code called SPEED~\citep[SPectral Elements in Elastodynamics with Discontinuous Galerkin,][]{Mazzieri_Stupazzini_Guidotti_Smerzini_2013}, developed at Politecnico di Milano, Italy. BB-SPEEDset contains up to 30 worldwide seismic scenarios of magnitude $M_W$ included between 3.4 and 7.4, occurred in the past. For each scenario, the authors employed SPEED to reproduce the earthquake physics and generate low-frequency waveforms $\xv\ptp$ up to a frequency varying between 1 and 3 Hz. The mesh refinement and geological configuration of each case-study determine such the frequency band of validity of the synthesized accelerograms. Moreover, \citet{Paolucci_et_al_2021} validated each and every scenario against recorded time histories at available recording stations. In doing so, o widen the frequency band of the yielded waveforms, and reach the 0-30 Hz range of interest for structural design, the authors applied ANN2BB~\citep{Paolucci_et_al_2018} to SPEED accelerograms. ANN2BB is a 2-layers MLP that predicts the recorded short period $Sa_{\yv[0]}$ spectral ordinates from the long period simulated ones $Sa_{\xv}$, obtained from SPEED time histories $\xv\ptp$. According to~\citet{Paolucci_et_al_2021}, BB-SPEEDset has no systematic bias in the trend of the generated waveforms, with respect to the NEar-Source Strong-motion version 2.0 (NESSv2) of recorded ground-motion dataset~\citep{Sgobba_Felicetta_Lanzano_Ramadan_DAmico_Pacor_2021}, which served as validation test bed for SPEED simulations.

\section{Results}
\label{sec-Results}
\begin{figure}
    \centering
    \subfloat[]{\includegraphics[width=0.45\textwidth]{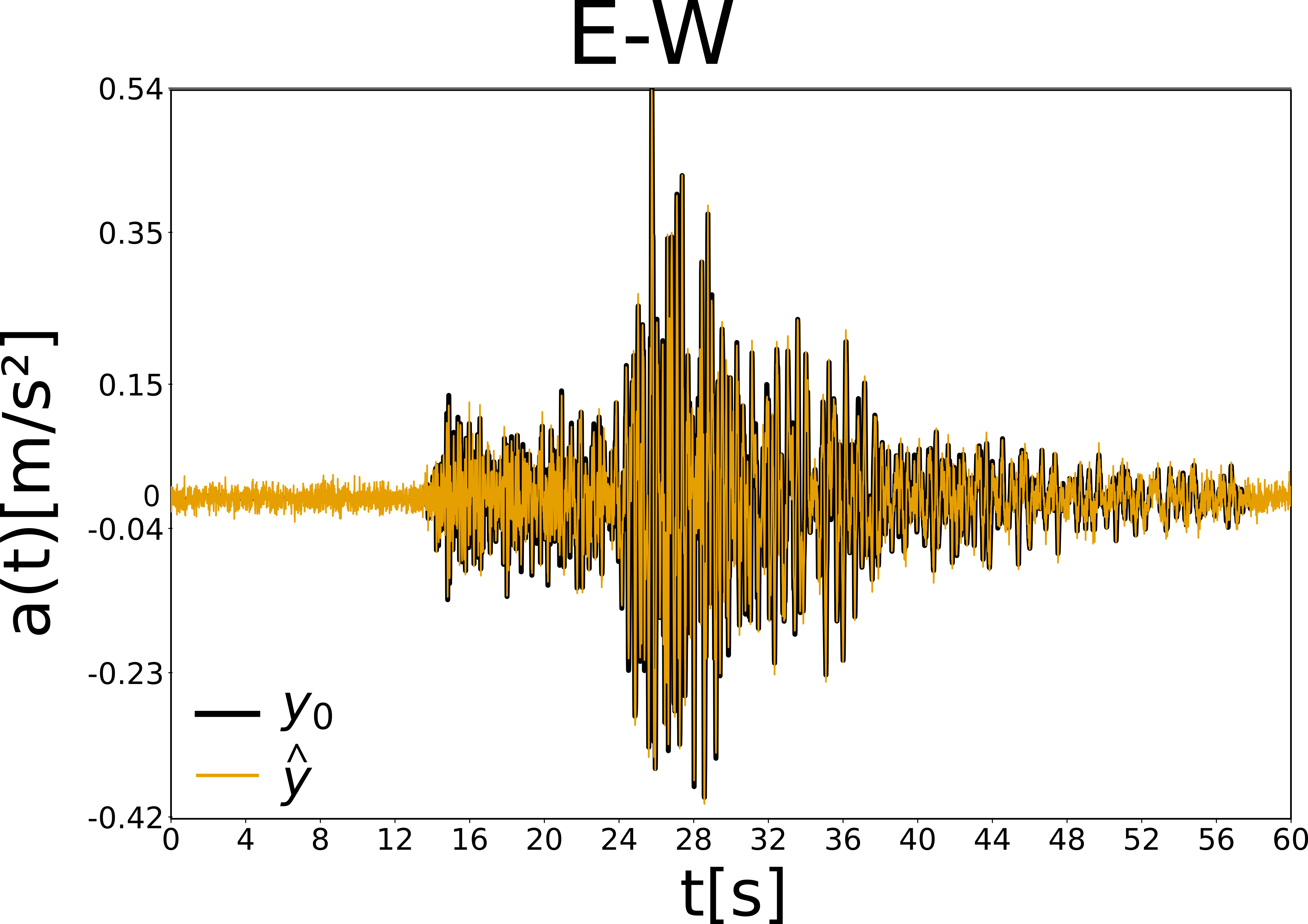}}
    \subfloat[]{\includegraphics[width=0.45\textwidth]{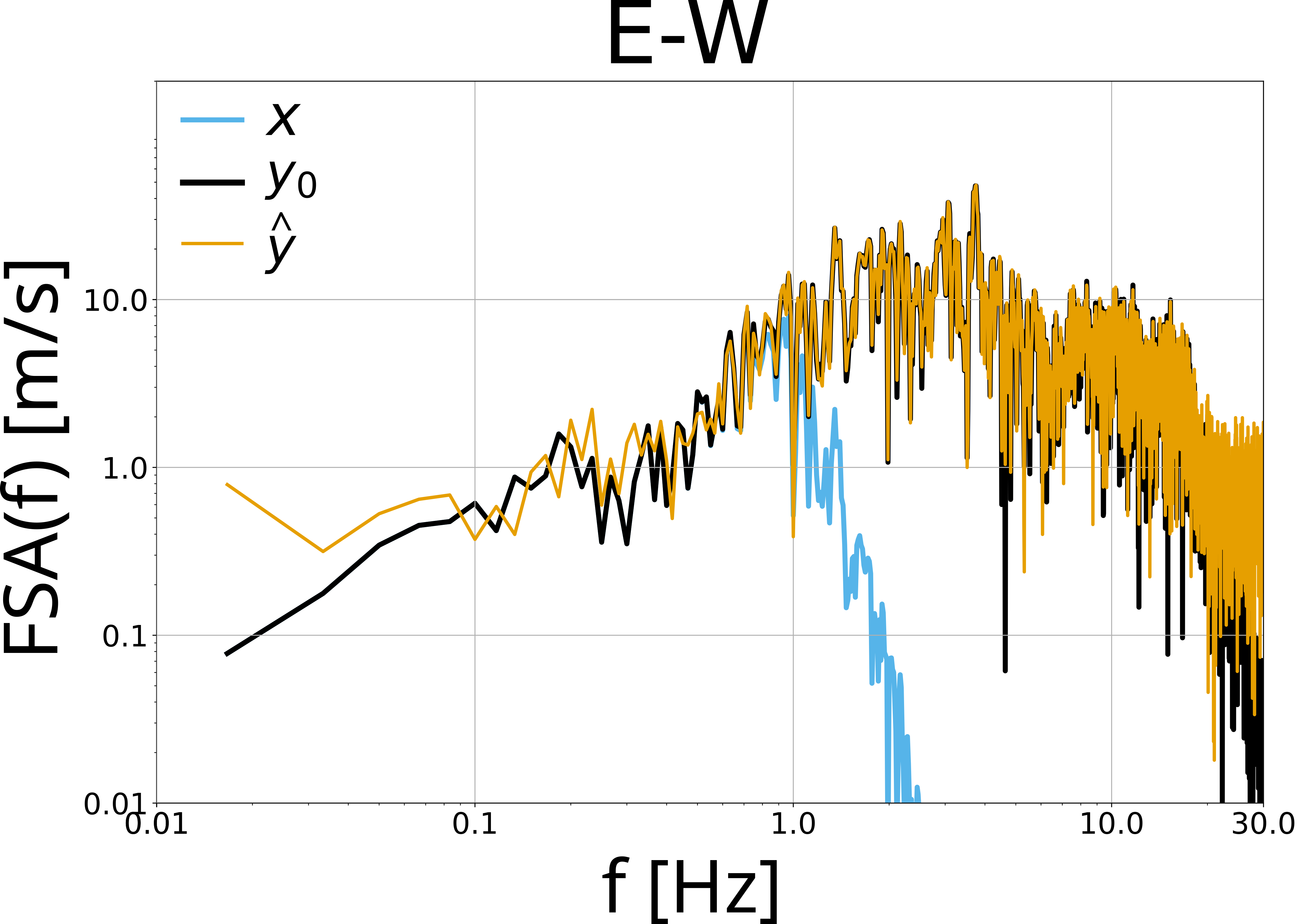}}\\
    \subfloat[]{\includegraphics[width=0.45\textwidth]{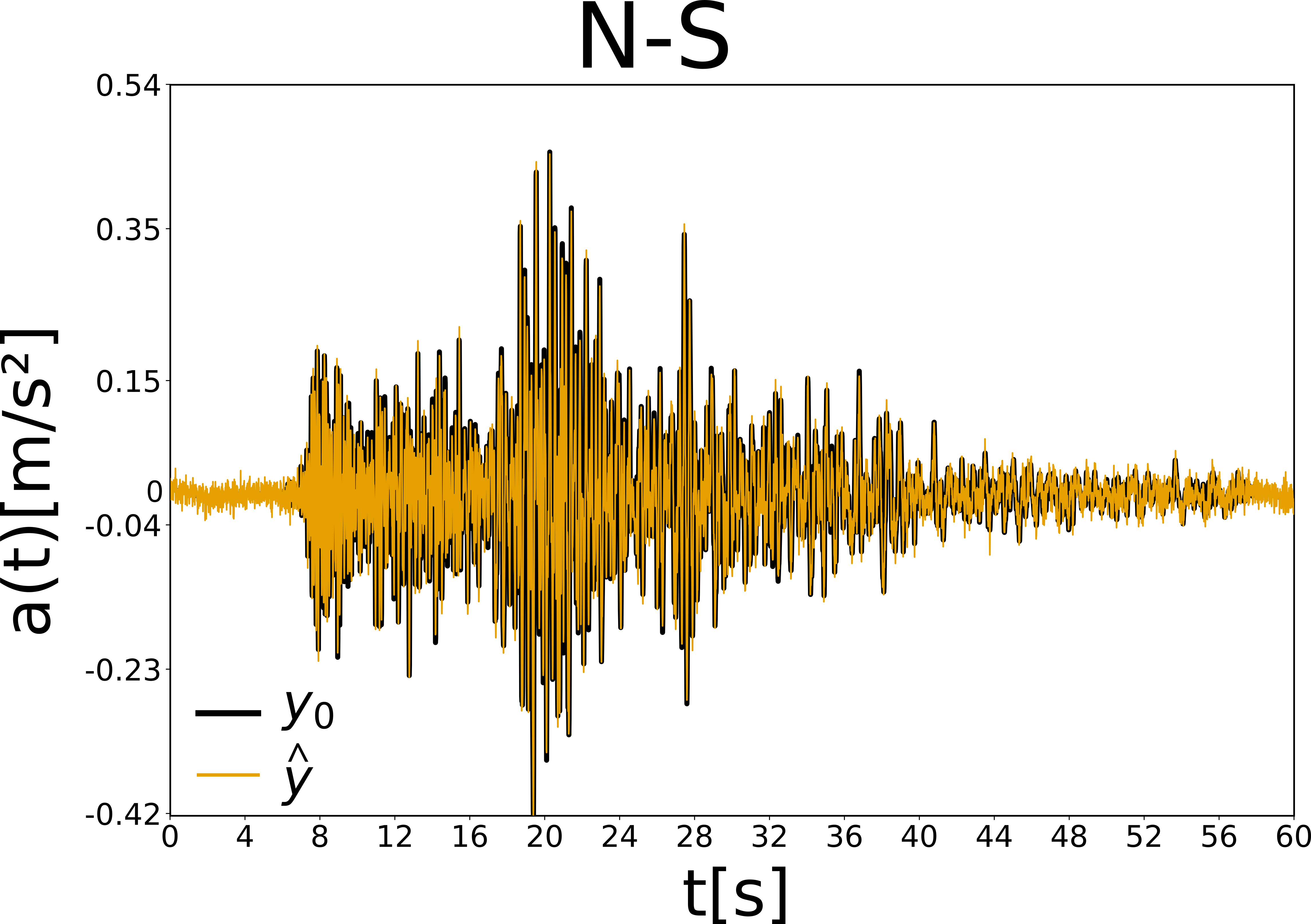}}
    \subfloat[]{\includegraphics[width=0.45\textwidth]{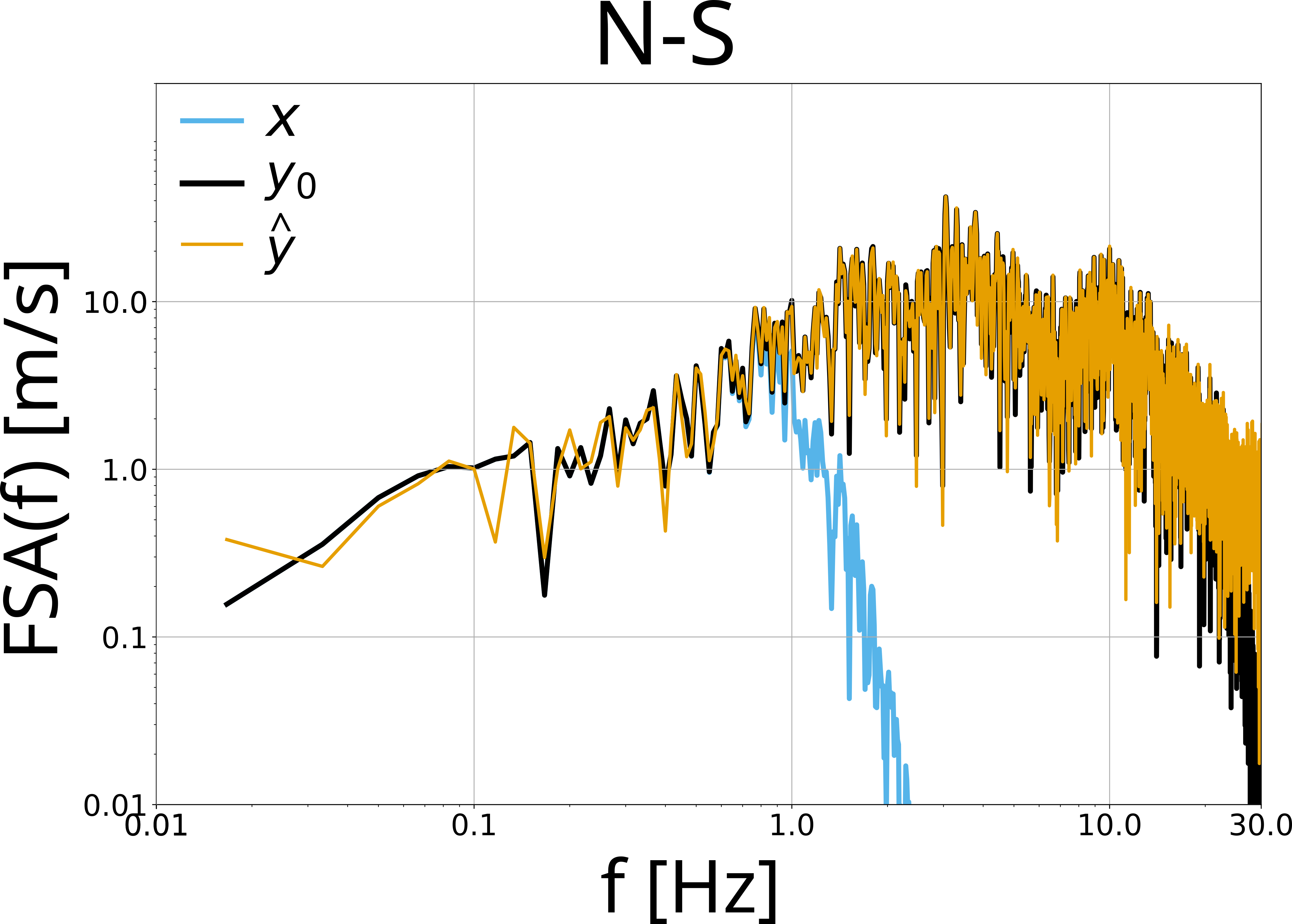}}\\
    \subfloat[]{\includegraphics[width=0.45\textwidth]{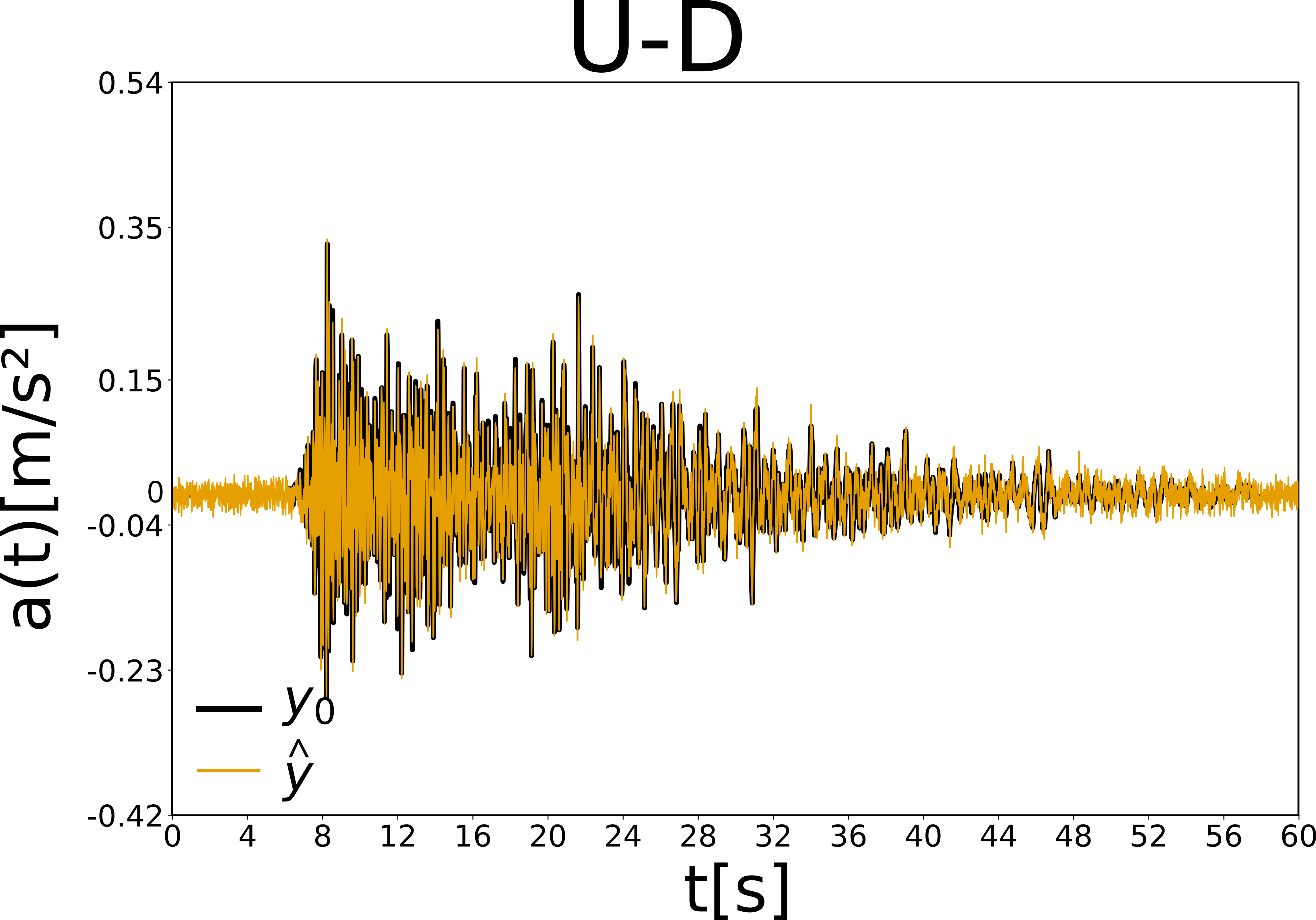}}
    \subfloat[]{\includegraphics[width=0.45\textwidth]{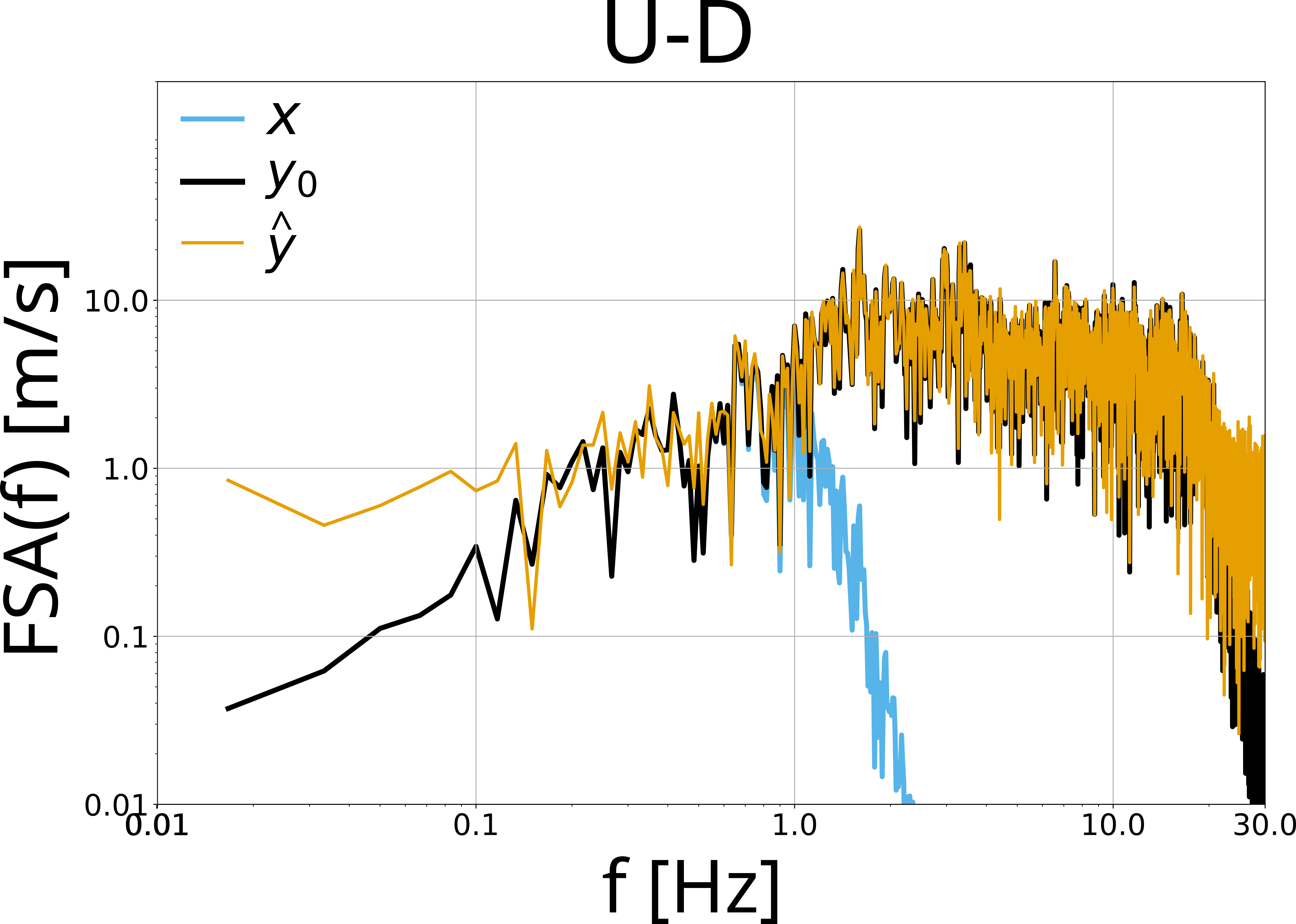}}
    \caption{EW (a), NS (c) and UD (e) time histories $\yv[0]$ ($a(t) [m/s^2]$, black) compared to DiT1D generation $\yhv$ (orange). EW (b), NS (d) and UD (f) Fourier's amplitude $FSA(f) [m/s]$, including $\xv[0]$.}
    \label{fig-amplitude_frequency_graph}
\end{figure}

As shown in \Cref{fig-amplitude_frequency_graph}, our combination of DiT1D and CNN-LSTM for amplitude prediction is capable to generate realistic broadband 3C seismic signals station-wise.
The MHCA layers are effectively guiding the generation, steering the low-to-high frequency extrapolation above 1 Hz. To assess DiT1D's performance, in \Cref{tab-results} we compare our model against several established baseline methods: a standard U-Net architecture, a diffusion-based denoising U-Net~\citep{Vivekananthan_2024}, and NU-Wave2~\citep{han_nu-wave_2022}, which corresponds to state-of-the-art diffusion model designed for neural audio up-sampling.
\begin{table}[ht!]
    \centering
    \begin{tabular}{lrrrrrr}
        \toprule
        Model & MSE & SNR & SSIM  & $D_s$  &  $\mathcal{T}_i$ [s]\\
        \midrule
        UNet & 3.3E$\phantom{}^{\text{-3}}$ & 12 & 0.87 & 2.5E$\phantom{}^{\mathbf{-3}}$ &  \textbf{0.002} \\
        UNet DDPM & 1.1$^{\text{-3}}$ & 14.2 & 0.89 & 2.7E$\phantom{}^{\text{-3}}$ &  6.5 \\
         DiT1D ($\Loss_{\text{ELBO}}$) & 4.1E$\phantom{}^{\text{-4}}$ & 17.8 & 0.94 & 5.1E$\phantom{}^{\text{-4}}$ & 0.71\\ 
        \textbf{DiT1D ($\Loss_{\text{ELBO}}+\lambda\mathcal{R}$)}  &\textbf{2.3E}$\phantom{}^{\mathbf{-4}}$&    \textbf{18.2} & \textbf{0.97} & \textbf{2.8E}$\phantom{}^{\mathbf{-5}}$ & 0.71 \\
        \bottomrule
    \end{tabular}
    \caption{Model testing metrics: Mean Square Error (MSE), Signal-to-Noise Ratio (SNR), Structural Similarity Index Measure (SSIM), $D_s$ and inference time $\mathcal{T}_i$.}
    \label{tab-results}
\end{table}
To ensure fair comparison, we downscaled all models to an equivalent amount of 150 million parameters. We employed several metrics such as Mean Square Error (MSE), Structural Similarity Index Measure (SSIM) and a spectral distance $D_s$ proposed by~\citet{zhang2022waveletknowledgedistillationefficient}. The latter corresponds to the mean relative distance between two signals $y_0$ and $\tilde{y}$ in the frequency domain, and it reads:
$$D_s\pdp{y_0,\tilde{y}} = \emean[f]\psp{\Big\vert\frac{\vert Y_0(f)\vert}{\Vert Y_0 \Vert_{L^2}}-\frac{\vert\tilde{Y}(f)\vert}{\Vert \tilde{Y} \Vert_{L^2}}\Big\vert}$$
with $Y_0(f)$, $\tilde{Y}(f)$ being the Fourier's transform of $y_0(t)$ and $\tilde{y}(t)$ respectively.
We compared the DiT1D trained with and without the penalty term $\mathcal{R}$ in the loss formulation in \Cref{eq-eq5bis}. \Cref{tab-results} proves that our DiT1D model outperforms all other models in all metrics, and that the penalty term is critical for its performance. Moreover, DDIM's generation time $\mathcal{T}_i$ remains comparable to non-diffusion approaches.

The comparative analysis presented in \Cref{tab-results} demonstrates that the proposed DiT1D architecture consistently outperforms both standard and advanced baseline methods across all evaluation metrics. These results validate the effectiveness of our approach for frequency augmentation in seismic signal processing. Notably, our analysis revealed an unexpected result: the classical U-Net 1D architecture outperformed more sophisticated approaches, suggesting that contemporary architectures may not be optimally suited for this specific signal processing task. NU-wave-2 yields impressive results for high frequency augmentation, but poorly performs in zero shot on earthquake datasets.
Moreover, the MSE of trained CNN-LSTM on the testing set was 0.04 m/s/s, meeting our performance requirements and providing realistic earthquake amplitudes when scaling normalized $\yhv$ generated by the DiT1D.

\subsection{Zero shot capability}
\label{subsec-Zero_shot_capability}
As introduced in \Cref{subsec-datasets_to_test_zero-shot_inference}, we tested the zero-shot capability of our model on two different datasets: the STEAD dataset and the BB-SPEEDset dataset. We employed the former database to test the DiT1D zero-shot performance in generating broadband time histories $\yhv\ptp$ based on the low-pass filtered version of the recorded one, namely $\xv[0]\ptp$. On the other hand, we exploited BB-SPEEDset to showcase the DiT1D capability of rendering broadband synthetic accelerograms, based on validated physics-based simulations $\xv$, reaching recording-level of realism. For the sake of comparison, we selected the 2009 $M_W$6.2 L'Aquila earthquake scenario (central Italy), focusing on AQK station~\citep[see ][for further details]{Smerzini_Villani_2012,Evangelista_et_al_2017,Paolucci_et_al_2021}. \citet{Smerzini_Villani_2012} and \citet{Evangelista_et_al_2017} constructed and refined the SPEED simulation of this earthquake scenario. The authors validated the SPEED simulation in the 0-2 Hz frequency range, comparing the synthetic time histories with the available recordings, at several stations nearby the epicenter. \citet{Evangelista_et_al_2017} reported an excellent fit of the synthetic and recorded time histories at AQK station, located within $a\approx$ 4 km from the epicenter. 

\begin{figure}[ht!]
     \centerline{\includegraphics[width=0.8\textwidth]{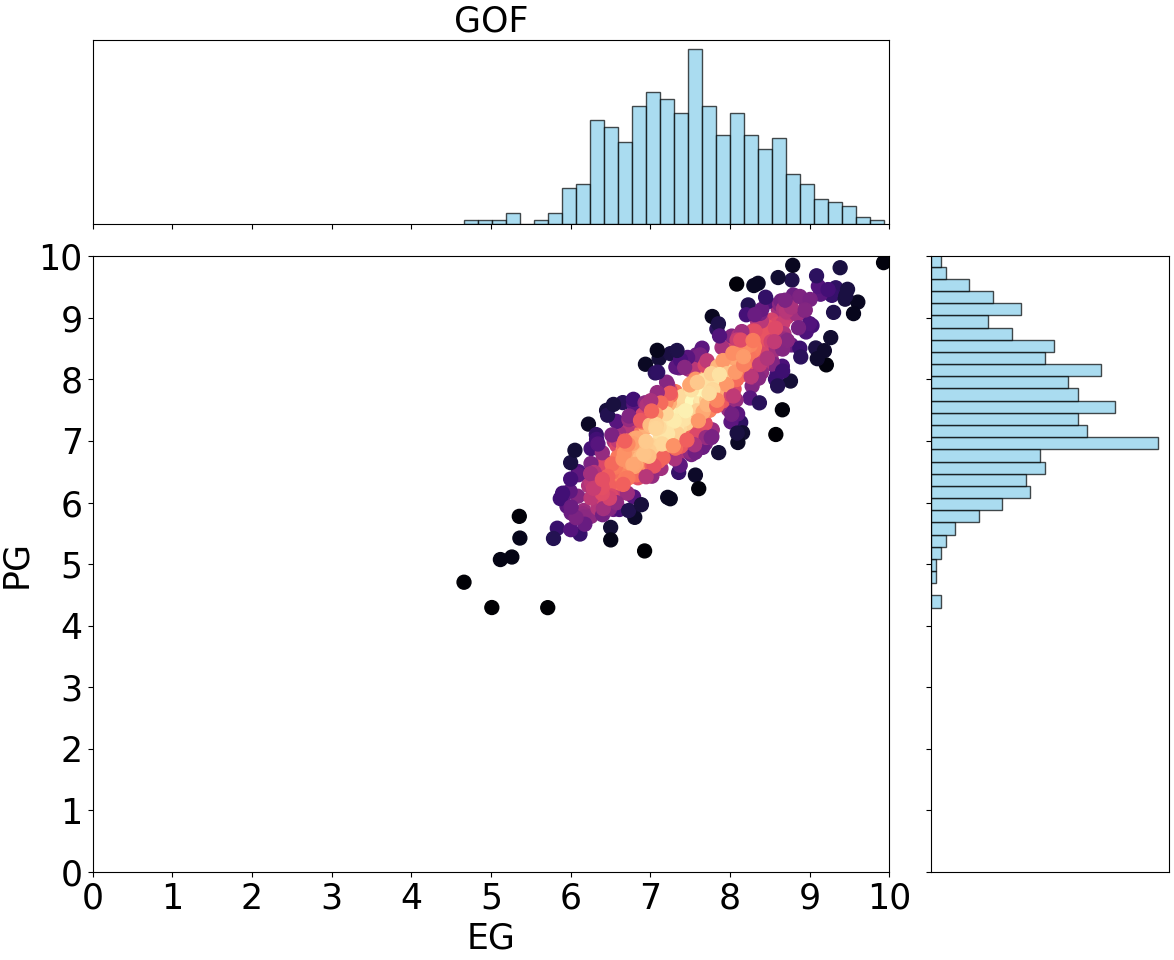}}
     \caption{$EG$ and $PG$ GOF values obtained on 500 random samples from STEAD database.}
     \label{fig-STEAD GOF graph}
\end{figure}

In order to fairly assess the performance of our model on the STEAD dataset, we adopted a Goodness-of-Fit (GOF) metrics widely adopted in seismology, and proposed by~\citet{Kristekova_Kristek_Moczo_2009}. Such GOF is expressed on a scale of 0 to 10, measuring the average time-frequency similarity between two accelerograms. The higher the GOF value, the more similar the two signals are. We assess a GOF score value for the envelope similarity ($EG$) and one for the phase similarity ($PG$). In practice, 0-4 is considered as a poor score, 4-6 as a fair score, 6-8 as a good, and 8-10 as excellent. \ref{sec-gof} explains in details how those metrics are constructed in ~\citet{Kristekova_Kristek_Moczo_2009} (see \Cref{eq-EG} and \Cref{eq-PG}). However, it must be noted that the in our analysis, the DiT1D is not conceived to reconstruct the entire input ground motion time histories, as in autoencoder architectures, but to generate as much diverse as possible signals, keeping fixed the low-frequency portion of the spectrum.  \Cref{fig-STEAD GOF graph} shows the GOF results obtained by applying DiT1D zero-shot on 500 earthquake accelerograms randomly selected from STEAD and low-pass filtered a 1 Hz. 
As shown in \Cref{fig-STEAD GOF graph}, the DiT1D yields a wide variety of signals with $EG$ and $PG$ values ranking from fair to excellent. In the former case, the generated accelerogram is very different from the original one, while in the latter case, the generated signal closely resembles the broadband recorded signal, solely based on its 0-1 Hz component. Overall, mean $EG$ and $PG$ scores range around 7, highlighting the importance of the guidance $\xv[0]$ provided to the DiT1D. However, the GOF dispersion between 5 and 10, certifies a good generalization capability of the model.

\begin{figure}[ht!]
    \centering
    \subfloat[]{\includegraphics[width=0.45\textwidth]{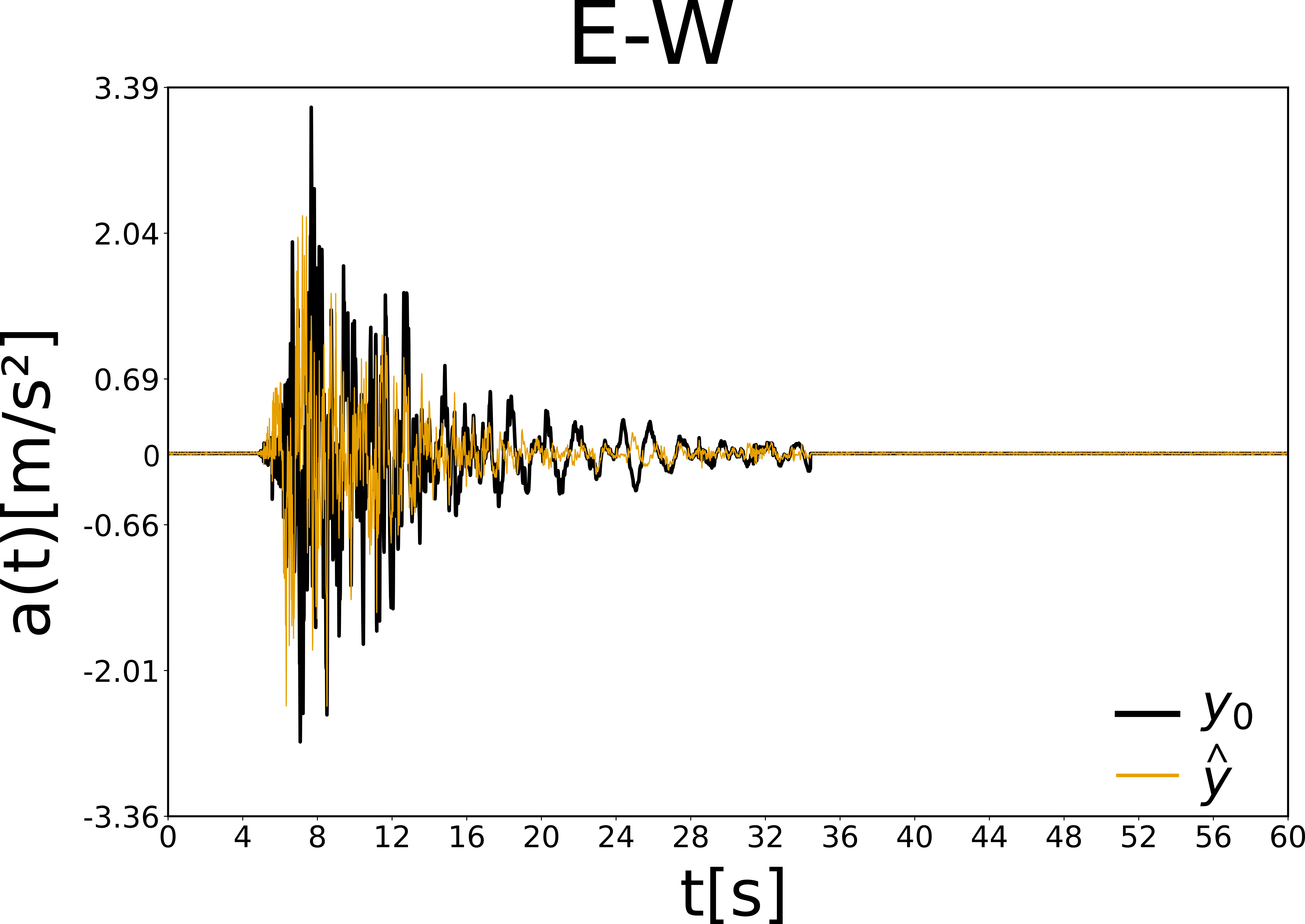}}
    \subfloat[]{\includegraphics[width=0.45\textwidth]{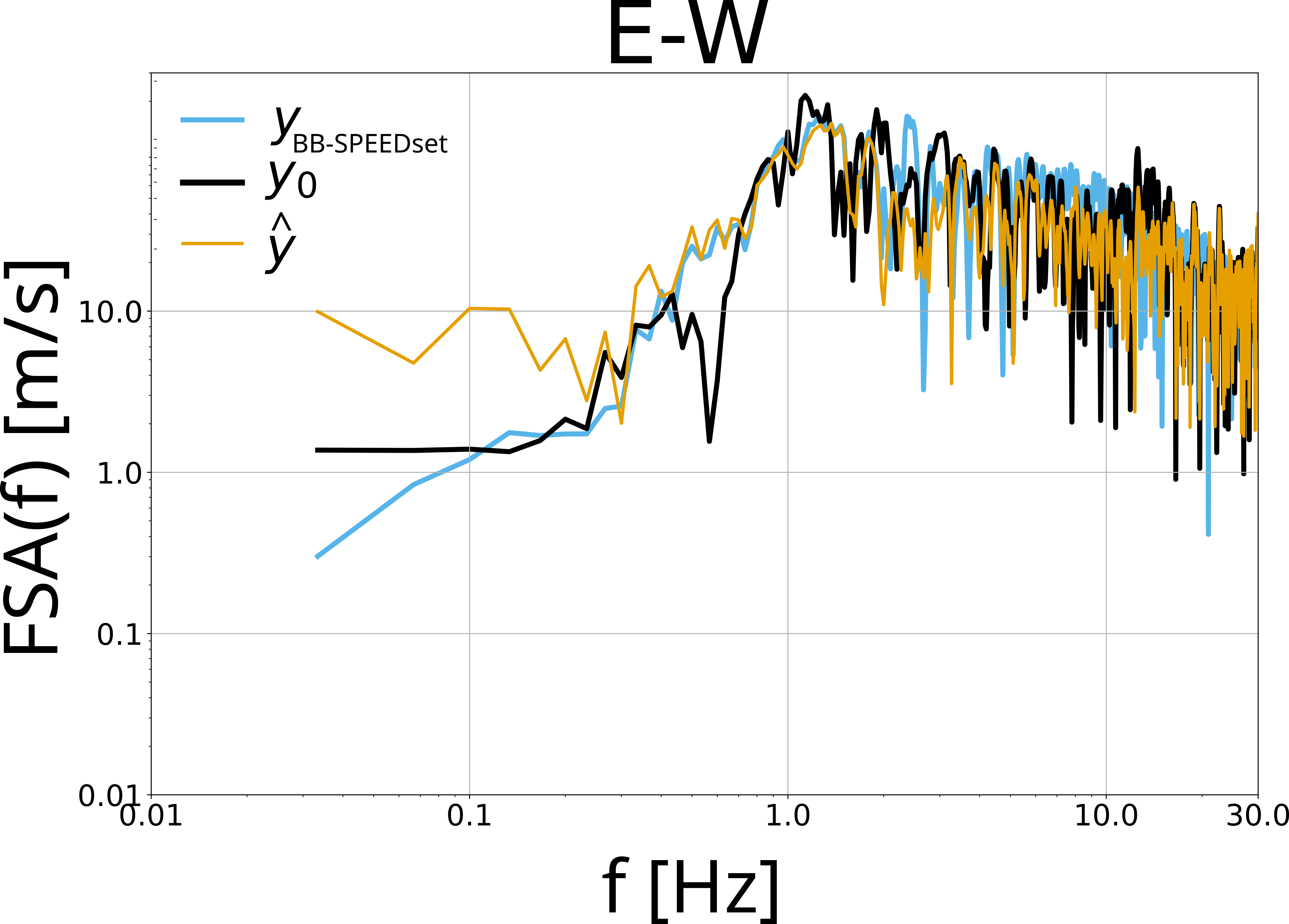}}\\

    \subfloat[]{\includegraphics[width=0.45\textwidth]{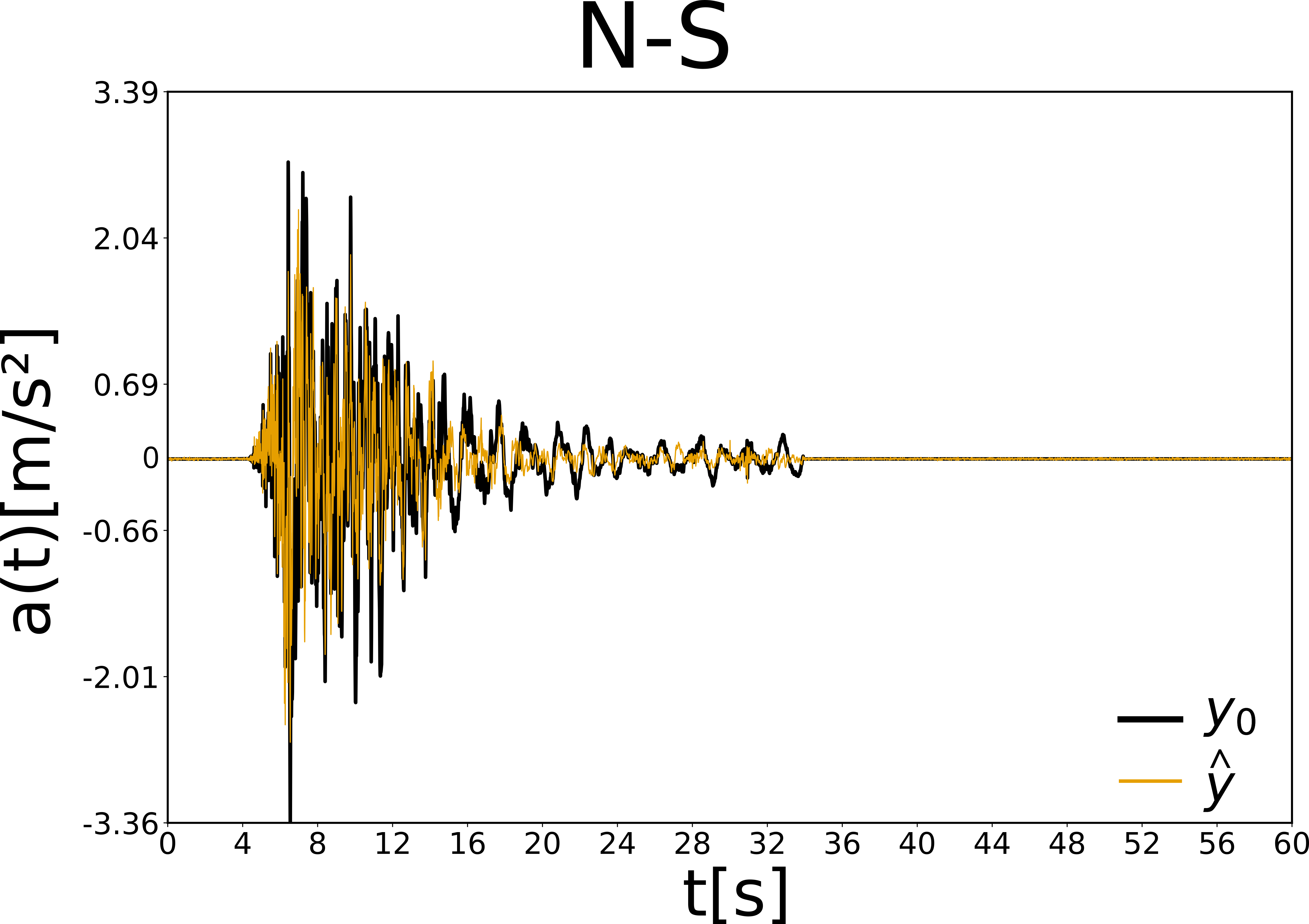}}
    \subfloat[]{\includegraphics[width=0.45\textwidth]{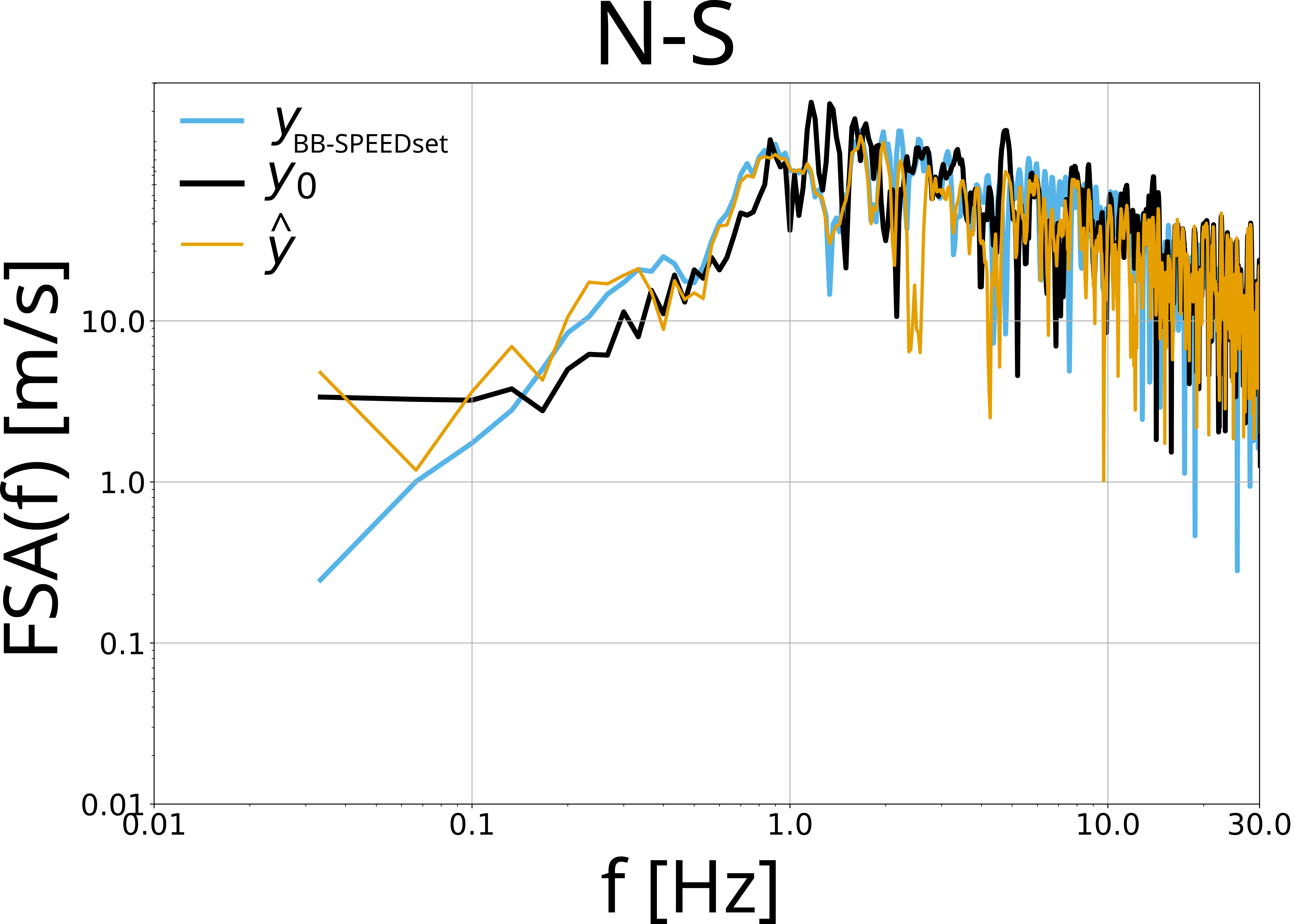}}\\
    
    \subfloat[]{\includegraphics[width=0.45\textwidth]{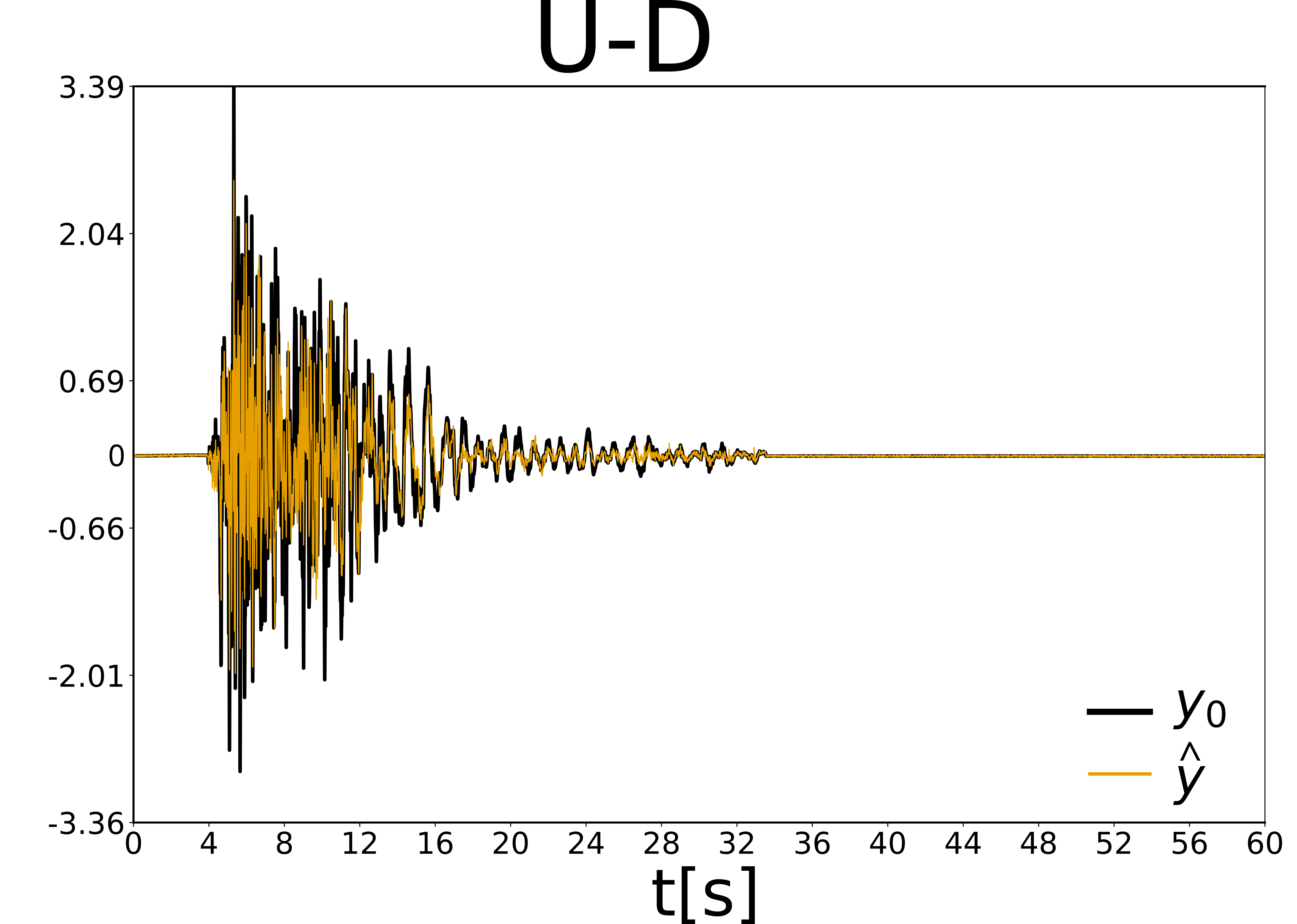}}
    \subfloat[]{\includegraphics[width=0.45\textwidth]{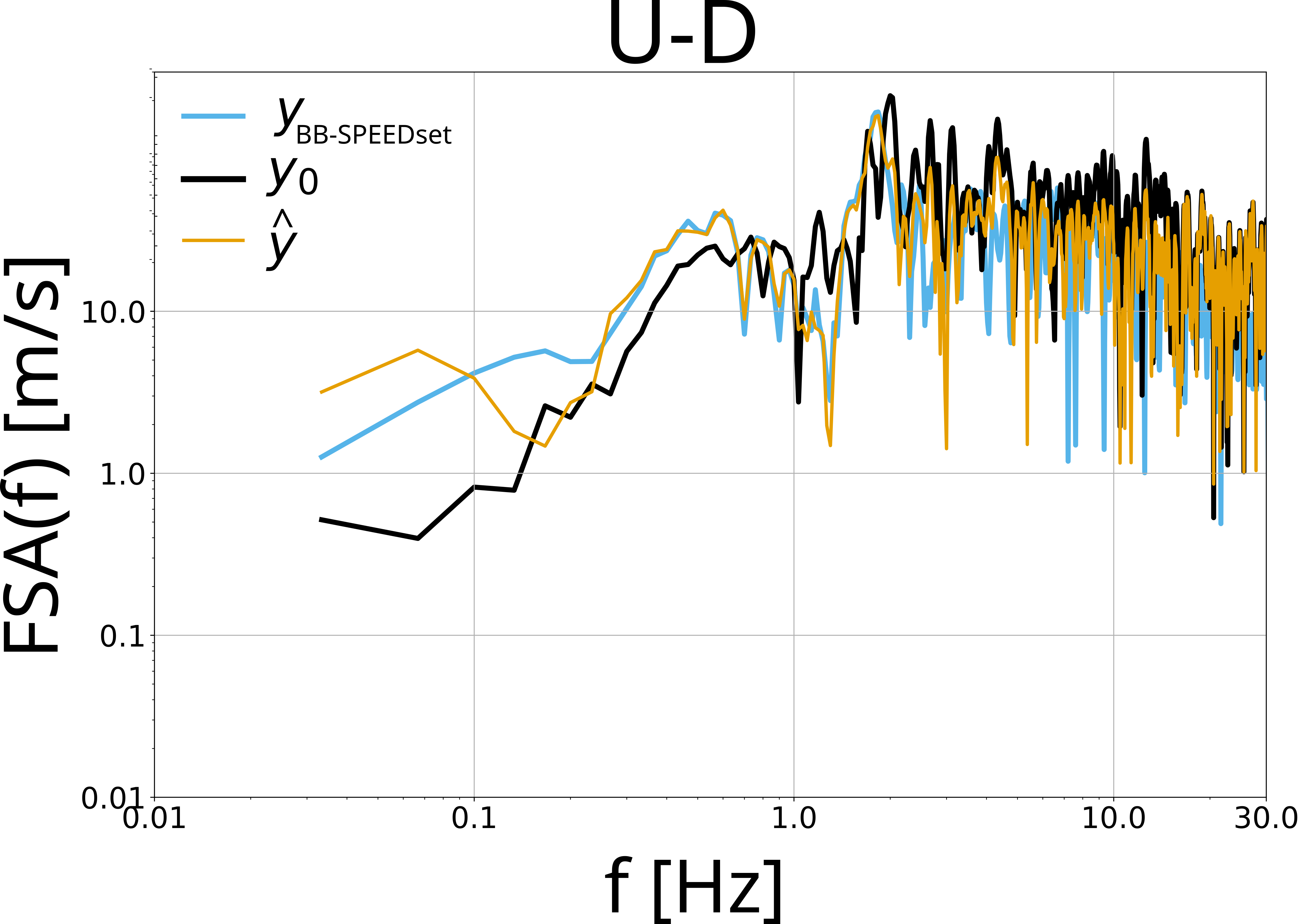}}
    \caption{EW (a), NS (c) and UD (e) time histories $\yv[0]$ ($a(t) [m/s^2]$, black) compared to DiT1D generation $\yhv$ (orange) at AQK station for the 2009 $M_W$6.2 L'Aquila earthquake. EW (b), NS (d) and UD (f) Fourier's amplitude $FSA(f) [m/s]$, including $\yv[\text{BB-SPEEDset}]$ from BB-SPEEDset simulations, low-pass filtered to 1 Hz.}
    \label{fig-AQK}
\end{figure}

In \Cref{fig-AQK}, we present the results of test performed on the 2009 $M_W$6.2 L'Aquila scenario, extracted from BB-SPEEDset database. Recorded $\yv[0]$ (NESSv2 database, black line), simulated $\yv[\text{BB-SPEEDset}]$ (corresponding to SPEED simulation between 0-2 Hz and ANN2BB from 2 Hz on, from BB-SPEEDset, blue line) and generated $\yhv$ (DiT1D, orange line) accelerograms at AQK station are reported, along with the respective Fourier's amplitude spectra. 
As shown in \Cref{fig-AQK}, the DiT1D outperforms the combination of SPEED and ANN2BB proposed by~\citet{Paolucci_et_al_2018}. Interestingly, we obtained the guidance signal $\xv$ by low-pass filtering the BB-SPEEDset time histories $\yv[\text{BB-SPEEDset}]$ at 1 Hz, despite the fact that the SPEED simulation are validated in the 0-2 Hz frequency range. We observe that our DiT1D better reproduces the high-frequency content of the recorded time history available in the NESSv2 database, despite the low frequency misfit between the SPEED simulation and the records. Compared to SPEED+ANN2BB, we improved the SSIM by 0.07. Feeding the 0-2 Hz low-pass-filtered simulated signals to the DiT1D does neither improve nor degrade significantly its performances. It must be noted that while ANN2BB introduces random phases at high frequency, following the original work of \citet{Sabetta_Pugliese_1996}, the DiT1D does not make this assumption, learning the low-to-high frequency mapping in both amplitude and phase.

\subsection{Computational Performance}
\label{subsec-Computational_Performance}
We used Jean Zay supercomputer of CNRS/IDRIS-GENCI to perform our experiments and draw performance measurements. Each computing node we used included:
%
\begin{itemize}

\item 2 CPUs Intel Xeon Gold 6346 at 3.1 GHz, with 16 hyperthreaded physical cores per processor and 1TB RAM for both processors.

\item 4 GPUs NVidia HGX A100, with 6912 Cuda cores, 432 $3^{rd}$ Tensor kernels and 40 GB RAM per GPU, and a $3^{rd}$ generation NVlink interconnecting the 4 GPUs at 2.4 TB/s.

\end{itemize}

The software we have developed and benchmarked is based on Python-3.10.12 and Pytorch-2.2.2 and some classical Python libraries like Numpy.

One epoch of the training dataset we used for our benchmark is compound of 14000 couples $\pdp{\xv[0],\yv[\tau]}$ of two time series of 6000 3D data points each (corresponding to 6000 $\times$3 4-bytes values per series): the conditioning $\xv[0]$ and the diffusion stage $\yv[\tau]$. However, inside each epoch we process data per batch of 32 couples. Each couple has 2$\times$ 6000 $\times$ 3 = 36000 4-bytes values, \textit{i.e.}, a size of 140.6 KBytes. Each batch of 32 couples has 32 $\times$ 36 000 = 1152000 4-bytes values, corresponding to a size of 4.4 MBytes. Finally, each epoch of 14000 couples requires to process 14000 $\times$ 36000 = 1152000 4-bytes values, for a total of 1.9 GBytes of data, and each epoch is compound of 14000 / 32 = 437.5 batches.

\begin{figure}
\centerline{
\includegraphics[width=10cm]{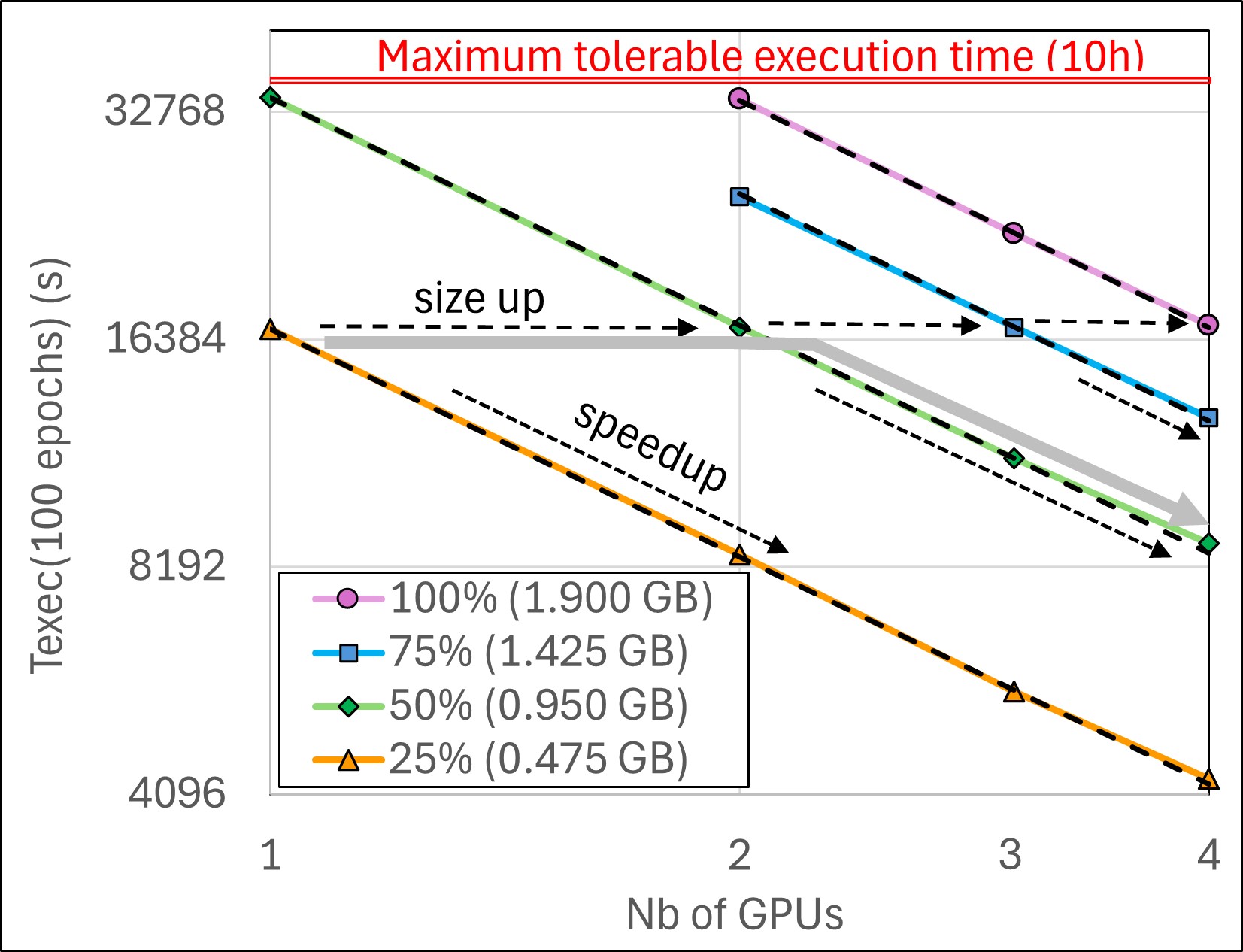}
}
\caption{\label{fig-ScalabilityCurves}Training times of 100 epochs for different dataset sizes on 1 to 4 GPUs NVidia HGX A100, inside one Jean Zay computing node and limited to 10h.}
\end{figure}

We measured the execution time of a complete training of 100 epochs of our generative models, from 1 up to 4 GPUs on one node of Jean Zay supercomputer, and for 4 different sizes of dataset: 25\%, 50\%, 75\% and 100\% of our 1.9 GBytes dataset. Ideal curves of execution time for a fixed dataset size and function of the number of GPU should correspond to the following expression:
\begin{equation}
\label{eq-perfectspeedup}
T_{exec}^{ideal}(n_{GPU}) = \frac{T_{exec}(1_{GPU})}{n_{GPU}}
\end{equation}
After measuring the training execution times on one GPU, we can draw these ideal execution time curves on \Cref{fig-ScalabilityCurves} in full logarithmic scales, in order to transform these hyperbolic curves in straight lines with a slope of $-1$. This graphical representation enables us to easily compare experimental curves with ideal curves, and to assess the \emph{speedup}, \emph{size up} and \emph{scalability} of our complete software and hardware system. We considered a maximum tolerable execution time of 10h (36000 s), and to limit our analysis to execution times below this limit.

\Cref{fig-ScalabilityCurves} shows the execution time curves we measured on our multi-GPU node for 4 increasing sizes of dataset.
The solid colored lines in \Cref{fig-ScalabilityCurves} are the measured experimental times for four fixed dataset sizes and the superposed dashed lines are the corresponding ideal time curves (\Cref{eq-perfectspeedup}). The experimental curves match the perfect curves and are parallel, which means that parallelization up to 4 GPUs presents negligible overheads for each tested dataset and produces perfect \emph{speedup}. Moreover, by doubling the dataset size we can maintain constant the computation time by doubling the number of GPUs, assuming we need the same number of epochs. This represents a perfect \emph{size up},  illustrated on \Cref{fig-ScalabilityCurves} with some horizontal dashed arrows. Processing larger datasets with more resources in the same time is important to avoid disturbing a working plan. Our hardware and software configuration this objective without overheads. As we obtained perfect speedup and perfect size up, we obtain a perfect \emph{scalability} on one multi-GPU Jean Zay node. This is comfortable when launching frequent trainings on different datasets: we can cumulate the speedup and size up approaches to satisfy some business constraints. For example, starting with one GPU to process the 25\% dataset, we can plan to use 4 GPUs to process the 50\% dataset 2 times faster if needed (following the grey arrow on \Cref{fig-ScalabilityCurves}). Then we can reserve the right number of computing resources for the right time on the supercomputer in advance.

\section{Discussion}
\label{sec-Discussion}
The DiT1D, in combination with the CNN-LSTM for amplitude estimation, proves to be a solid and reliable tool to generate realistic broadband 3C seismic signals. Compared to other state-of-the-art models, it is capable of generating realistic broadband signals, preserving the low-frequency part of the spectrum, encoding the minimum knowledge on the physics of the earthquake. This means that most of the features that empirical Ground Motion Models struggle to adequately reproduce are captured by the DiT1D. Site-effects, directivity, spatial correlation and many other peculiarities of a ground shaking event can be modelled with a standard numerical solver and fed as input to the DiT1D, which is agnostic to the employed software. Since we have not adopted any cumbersome spatial attention layer, our solution represents a reliable and lightweight alternative to station-wise improvement of the numerical model. Due to its station-wise nature, the DiT1D can be deployed on simulations with varying spatial resolution, privileging low-to-medium-size models. The latter are not only cheap to run, but also easier to constraint with sound physical assumptions and parameter estimation. The DiT1D generates the missing medium-high frequency portion of the spectrum, reaching a recording level of realism. We proved that not only this is true when guiding the DiT1D with low-pass filtered signals, but also when feeding validated numerical simulations to it (such as the BB-SPEEDset synthetics). Interestingly, in \Cref{subsec-Zero_shot_capability}, we showed that even when masking part of the frequency content of the conditioning signal $\xv$ to the DiT1D, it is capable to realistically infer the high-frequency spectrum.

From a computational perspective, the DiT1D training process perfectly scales across multiple GPUs, opening the door to portable fine-tuning post-training. Moreover,\Cref{tab-results} reveals that the DiT1D is capable of generating realistic signals in a fraction of the time required by other state-of-the-art models. Therefore, the DiT1D represents a state-of-the-art AI generative surrogate model for strong ground motion prediction that one can employ off-the-shelf. 

Despite the above-mentioned advantages, the DiT1D struggles to render multiple broadband signals out of the same conditioning $\xv$, especially when we penalize the training loss with the L2 norm of the high-frequency residual $\mathcal{R}$. To unlock one-to-many generative capability, we would require a dedicated database and a different training strategy. As a matter of fact, since for any recording $\yv[0]$ there is only one corresponding low-pass filtered signal $\xv[0]$, one-to-many mapping could be learned by running several low-frequency simulations corresponding to the same recording, and employing the ELBO loss in \Cref{eq-eq5}. To reduce the computation burden, \citet{Perrone_Lehmann_Gabrielidis_Fresca_Gatti_2025} proposed to adopt a neural operator surrogate to quickly sample realistic physics-based guiding time histories $\xv$ to be fed to the DiT1D. We leave this as future work. 

Finally, \Cref{fig-AQK} highlights the fact that the DiT1D may slightly deviate from the conditioning signal $\xv\ptp$ at low frequency. This aspect has to do with the effectiveness of the MHCA layers, but also with the denoising process, which, depending on how many denoising steps are carried out, might leave some spurious low-frequency residuals. This aspect has been highlighted by~\citet{Gatti_Clouteau_2020} for GANs, and is partially due to the fact that we did not enforce any baseline correction at training time, to avoid over-constraining the generative process. From the time history plots in \Cref{fig-AQK}, we observe that this discrepancy does not alter the realism of the generated samples $\yhv$.
\section{Conclusions}
\label{sec-Conclusions}
In this paper, we proved the viability of the transformer architecture as powerful generative tool of broadband seismic signals. We employed a transformer architecture, duly adapted to 3C time histories, and used it a denoising diffusion backbone network, for conditional generation of seismic signals. We guide the diffusion transformer with low frequency signal conditioning, relying on low-pass-filtered records at training time, and on numerical simulations at inference time. Our approach simplifies the training process compared to existing methods while reaching superior generative performance on different databases. Moreover, the DiT1D outperforms the state-of-the-art ANN2BB technique when enhancing numerically simulated earthquake ground motion accelerograms. 

These encouraging results pave the way to the routinary use of the DiT1D as an off-the-shelf surrogate model to enhance the realism physics-based earthquake simulations at regional scale.
In future work, we will focus on enhancing both the efficiency and versatility of the generation process. As our goal is to generate diverse yet realistic seismic events, we aim to explore one-to-many generation techniques, where multiple high-frequency signals are derived from a single low-frequency guiding input. 

\section*{Acknowledgements}
The DiT1D and accessory CNN-LSTM were trained on the A100 GPU nodes available at Ruche, the Université Paris-Saclay supercomputer facility. Moreover, this work was granted access to the HPC resources of IDRIS under the allocation 2024-A0170414346 and 2025-AD011015929 made by GENCI. 
This study is part of the MINERVE project with aims to develop innovative digital methods for railway infrastructures. The MINERVE project has been financed by the French government within the framework of France 2030.

The authors would like to thank Dr. Chiara Smerzini, Dr. Manuela Vanini and Prof. Roberto Paolucci from Politecnico di Milano, for their valuable guidance on the use of BB-SPEEDset and NESS dataset, as well as for providing the output of SPEED simulations adopted to validate BB-SPEEDset.
\appendix
\section{\citet{Kristekova_Kristek_Moczo_2009}'s Goodness-of-Fit (GOF) metrics}
\label{sec-gof}
The \citet{Kristekova_Kristek_Moczo_2009}'s GOF are time--frequency metrics, based on the continuous wavelet transform (CWT) $ L^2(\mathbb{R})\ni s \mapsto W\in L^2(\R\times \R^{+}_{*})$ of the time history $s(t)$, that reads:
\begin{equation}
    W\psp{s}(t,f) = \sqrt{\frac{2\pi \vert f \vert }{\omega_0}} \int_{\R} s(\tau)\psi^{*} \left(2\pi f\frac{\tau-t}{\omega_0}\right) d\tau
    \label{eq-cwt}
\end{equation}
where $\psi$ is the chosen wavelet basis, $t$ represents its time translation  and $\omega_0$ its scale parameter. $*$ represents the complex conjugate operation and $f$ the frequency value. In~\citet{Kristekova_Kristek_Moczo_2009} $\psi$ is chosen as the Morlet's wavelet, which reads:
\begin{equation}
\psi (t) = \pi^{-1/4}\text{e}^{i\omega_0t}\text{e}^{-t^2/2}
\label{eq-morlet}
\end{equation}
with $\omega_0$ = 6~\citep{Kristekova_Kristek_Moczo_2009,Mallat_2009}.
The CWT in \Cref{eq-cwt} is a time-frequency ($TF$) representation of the signal $s\ptp$. Its envelope $A(t,f)$ and phase $\phi(t,f)$ at a specific point on the $TF$ plane read:
\begin{equation}
    A(t,f) = \left \vert W(t,f)\right \vert  \qquad \phi(t,f) = \text{Arg}[W(t,f)].
\end{equation}

Considering a signal $s\ptp$ and a reference one $s_r\ptp$, the metric proposed by~\citet{Kristekova_Kristek_Moczo_2009} compares the two signals based on the envelope misfit $\Delta A(t,f)$ and the phase one $\Delta \phi(t,f)$ which read respectively:
\begin{equation}
\Delta A(t,f) = A(t,f) - A_r(t,f) = \left \vert W(t,f)\right \vert  - \left \vert W_r(t,f)\right \vert
\end{equation}

\begin{equation}
\Delta \phi(t,f) = \phi(t,f) - \phi_r(t,f) = \text{Arg}[W(t,f)] - \text{Arg}[W_r(t,f)] = \text{Arg}\left[\frac{W(t,f)}{W_r(t,f)}\right]
\end{equation}
The two $TF$ misfits are normalized according to the following expressions, to obtain $TF$ envelope misfit:
\begin{equation}
    TFEM(t,f)=\frac{\Delta A(t,f)}{A_{\text{ref}}(t,f)}
\end{equation}
and $TF$ phase misfit $TFPM(t,f)$
\begin{equation}
TFPM(t,f) = \frac{\Delta \phi(t,f)}{\pi}
\end{equation}
Following \citet{Kristekova_Kristek_Moczo_2009}, both $TFEM$ and $TFPM$ metrics are weighted in time-frequency by their corresponding CWT of $s\ptp$, so to obtain time envelope/phase misfit (referred as to $TEM$ and $TPM$) and frequency envelope/phase misfit (\textit{i.e.}, $FEM$ and $FPM$ respectively) respectively, which read:
\begin{equation}
TEM(t) = \frac{\sum_{f} \left \vert W(t,f)\right \vert  \text{TFEM}(t,f)}{\left \vert W(t,f)\right \vert } 
\end{equation}

\begin{equation}
TPM(t) = \frac{\sum_{f} \left \vert W(t,f)\right \vert  \text{TFPM}(t,f)}{\left \vert W(t,f)\right \vert }
\end{equation}

\begin{equation}
FEM(f) = \frac{\sum_{t} \left \vert W(t,f)\right \vert  \text{TFEM}(t,f)}{\left \vert W(t,f)\right \vert }
\end{equation}

\begin{equation}
FPM(f) = \frac{\sum_{t} \left \vert W(t,f)\right \vert  \text{TFPM}(t,f)}{\left \vert W(t,f)\right \vert }
\end{equation}

Finally, the normalized single--valued envelope (EM) and phase (PM) misfits are obtained according to the following expressions:
\begin{equation}
EM = \sqrt{\frac{\sum_{f} \sum_{t} \left \vert W_r(t,f)\right \vert ^2 \left \vert \text{TFEM}(t,f)\right \vert ^2}{\sum_{f} \sum_{t}\left \vert W_r(t,f)\right \vert ^2}}
\end{equation}

\begin{equation}
PM = \sqrt{\frac{\sum_{f} \sum_{t} \left \vert W_r(t,f)\right \vert ^2 \left \vert \text{TFPM}(t,f)\right \vert ^2}{\sum_{f} \sum_{t}\left \vert W_r(t,f)\right \vert ^2}}.
\end{equation}
The GOF shown in \Cref{fig-STEAD GOF graph} are obtained by computing the Time-Frequency Envelope Goodness ($TFEG$), Time Envelope Goodness ($TEG$), Frequency Envelope Goodness ($FEG$) and Envelope Goodness ($EG$), computed as:
\begin{equation}
TFEG(t,f) = 10\cdot\text{e}^{-\left \vert \text{TFEM}(t,f)\right \vert}
\end{equation}
\begin{equation}
TEG(t) = 10\cdot\text{e}^{-\left \vert \text{TEM}(t)\right \vert}
\end{equation}
\begin{equation}
FEG(f) = 10\cdot\text{e}^{-\left \vert \text{FEM}(f)\right \vert}
\end{equation}
\begin{equation}
EG = 10\cdot\text{e}^{-\left \vert \text{EM}\right \vert}
\label{eq-EG}
\end{equation}

Equivalently, for the phase:
\begin{equation}
TFPG(t,f) = 10\cdot\text{e}^{1-\left \vert \text{TFPM}(t,f)\right \vert}
\end{equation}
\begin{equation}
TPG(t) = 10\cdot(1-\left \vert \text{TPM}(t)\right \vert)
\end{equation}
\begin{equation}
FPG(f) = 10\cdot(1-\left \vert \text{FPM}(f)\right \vert)
\end{equation}
\begin{equation}
PG = 10\cdot(1-\left \vert \text{PM}\right \vert)
\label{eq-PG}
\end{equation}

\bibliographystyle{plainnat}
\bibliography{main}

\end{document}